\input epsf

\documentstyle[twocolumn]{mn}

\def\beq{\begin{equation}}
\def\eeq{\end{equation}}
\def\bey{\begin{eqnarray}}
\def\eey{\end{eqnarray}}
\def\RM{\rm}
\def\d{\rm d}

\title[]
{The microlensing rate and mass function vs. dynamics of the Galactic bar}
\author[]
{HongSheng Zhao \& P.\ Tim de Zeeuw
\thanks{E-mail: hsz@strw.leidenuniv.nl, tim@strw.leidenuniv.nl} \\
Sterrewacht Leiden, Niels Bohrweg 2, 2333 CA, Leiden, The Netherlands
\thanks{Based in part on work done at the Max-Planck-Institut 
        f\"ur Astrophysik, Karl-Schwarzschild-Strasse 1,
        85740 Garching, Germany}}
\date{Accepted $\dots$ 
      Received $\ldots$;
      in original form $\ldots$ }
\pagerange{\pageref{firstpage}--\pageref{lastpage}}
\pubyear{1998}

\begin{document}
\maketitle
\label{firstpage}

\begin{abstract}

\noindent
With the steady increase of the sample size of observed microlenses
towards the central regions of the Galaxy, the main source of the
uncertainty in the lens mass will shift from the simple Poisson noise
to the intrinsic non-uniqueness of our dynamical models of the inner
Galaxy, particularly, the Galactic bar.  We use a set of simple
self-consistent bar models to investigate how the microlensing event
rate varies as a function of axis ratio, bar angle and velocity
distribution. The non-uniqueness of the velocity distribution of the
bar model adds a significant uncertainty (by about a factor of 1.5) to
any prediction of the lens mass.  Kinematic data and self-consistent
models are critical to lift the non-uniqueness.  We discuss the
implications of these results for the interpretation of microlensing
observations of the Galactic bulge.  In particular we show that
Freeman bar models scaled to the mass of the Galactic bulge/bar imply
a typical lens mass of around $0.8M_\odot$, a factor of 3-5 times
larger than the value from other models.  
\end{abstract}


\section{Introduction}

Five years of searches for gravitational microlenses, which
significantly increase the brightness of a background source star when
they come to the same line of sight by chance, have produced roughly
200 possible microlensing events towards the Galactic bulge/bar from
several surveys (e.g., Udalski et al.\ 1994; Alard et al.\ 1995;
Alcock et al.\ 1997; The expectation is that with more and more events
coming in at a steady rate, we will obtain a better understanding of
the structure of the Galaxy, and of the mass spectrum of lenses (see
Pacyz\'nski 1996 for a review).  Microlensing is perhaps the only
direct method to probe the mass function in the Galactic bulge/bar, in
particular the fraction of low luminosity objects (brown
dwarfs/M-dwarfs) just below or just above the hydrogen burning
limit. These objects are so faint intrinsically ($M_V>10$mag) and so
far away (distance modulus of about 14.5 mag) that they are missing in
even the deepest star count studies of the bulge/bar with the Hubble
Space Telescope (Gould, Bahcall \& Flynn 1996).

Several attempts have been made to estimate the typical mass of the
observed microlenses (Kiraga \& Paczy\'nski 1994; Zhao, Spergel \&
Rich 1994; Han \& Gould 1996; Mao \& Paczy\'nski 1996) on the basis
that events last longer for more massive lenses, with $m \propto t^2$,
where $m$ is the mass of a single point lens and $t$ is the time for a
source on the Einstein ring (where the amplification factor is 1.34)
to move to exactly behind the lens.  The time scale $t$ can be derived
by fitting the light curve of the amplified source, but to convert it
to the lens mass $m$ requires knowing other system parameters,
including the distances and transverse velocities of the lens and the
source.  These, unfortunately, are often poorly known from
observations.  Nevertheless the problem can be partially circumvented
if one has a well-determined dynamical model of the inner Galaxy.  The
missing information can then be simulated by drawing random samples of
the lens and the source in a Monte-Carlo fashion from the model
phase-space distribution of the inner Galaxy.

With the increase of the event sample size, it becomes meaningful to
ask the question whether the whole mass spectrum of the lenses can be
determined in the above way (Han \& Lee 1997).  This requires an
understanding of the uncertainty of the underlying dynamical model.
Models of the inner Galaxy are, unfortunately, still far from being
well-determined and they are subject to constant modifications and
improvements, driven by new observations (see e.g., reviews by de
Zeeuw 1993; Gerhard 1996).  How much the Galactic bar differs from a
simple oblate rotator is an unsettled issue.  Generally speaking the
density distribution of the Galactic bar is constrained only up to a
one-parameter sequence by the integrated COBE map, with the bar angle
being a free parameter (Binney, Gerhard \& Spergel 1997; Zhao 1997).
Various velocity structures of the bar could also be consistent with
the same bar potential (Pfenniger 1984).  Schwarzschild-type models
for the inner Galaxy (Zhao 1996; Binney et al.\ 1997) which attempt to
fit simultaneously the photometric and kinematic data measurements are
still rare.

The observed velocity field of a bar is the result of its tumbling
motion, the internal streaming motion on top of the pattern rotation,
and the velocity anisotropy.  Bars of different pattern speed and/or
orbital compositions are indistinguishable in their projected density
and optical depth maps, but they differ at the level of event rate and
event duration distribution.
%
\begin{figure} 
\epsfysize=6cm 
\centerline{\epsfbox{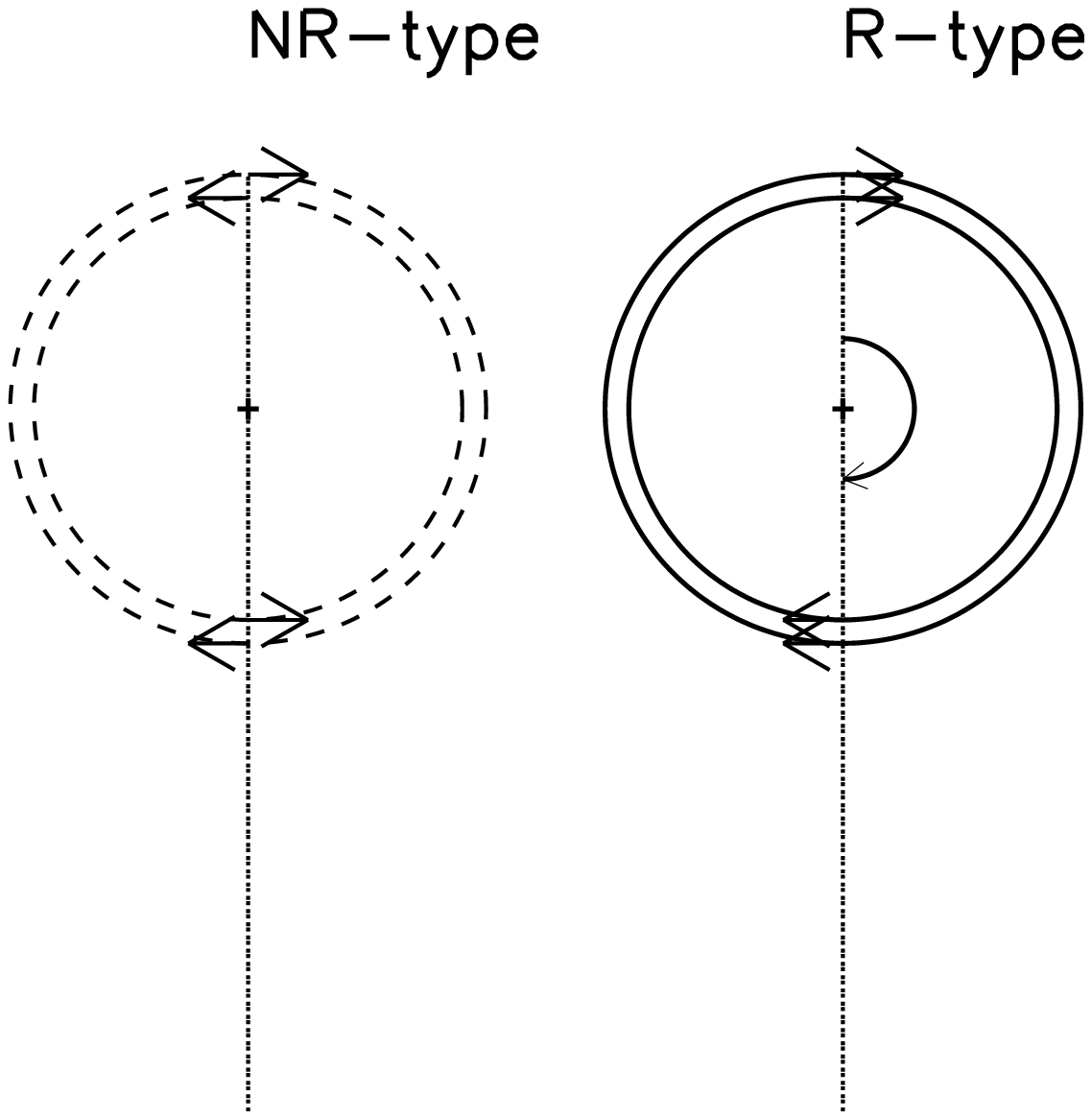}}
\epsfysize=4cm 
\centerline{\epsfbox{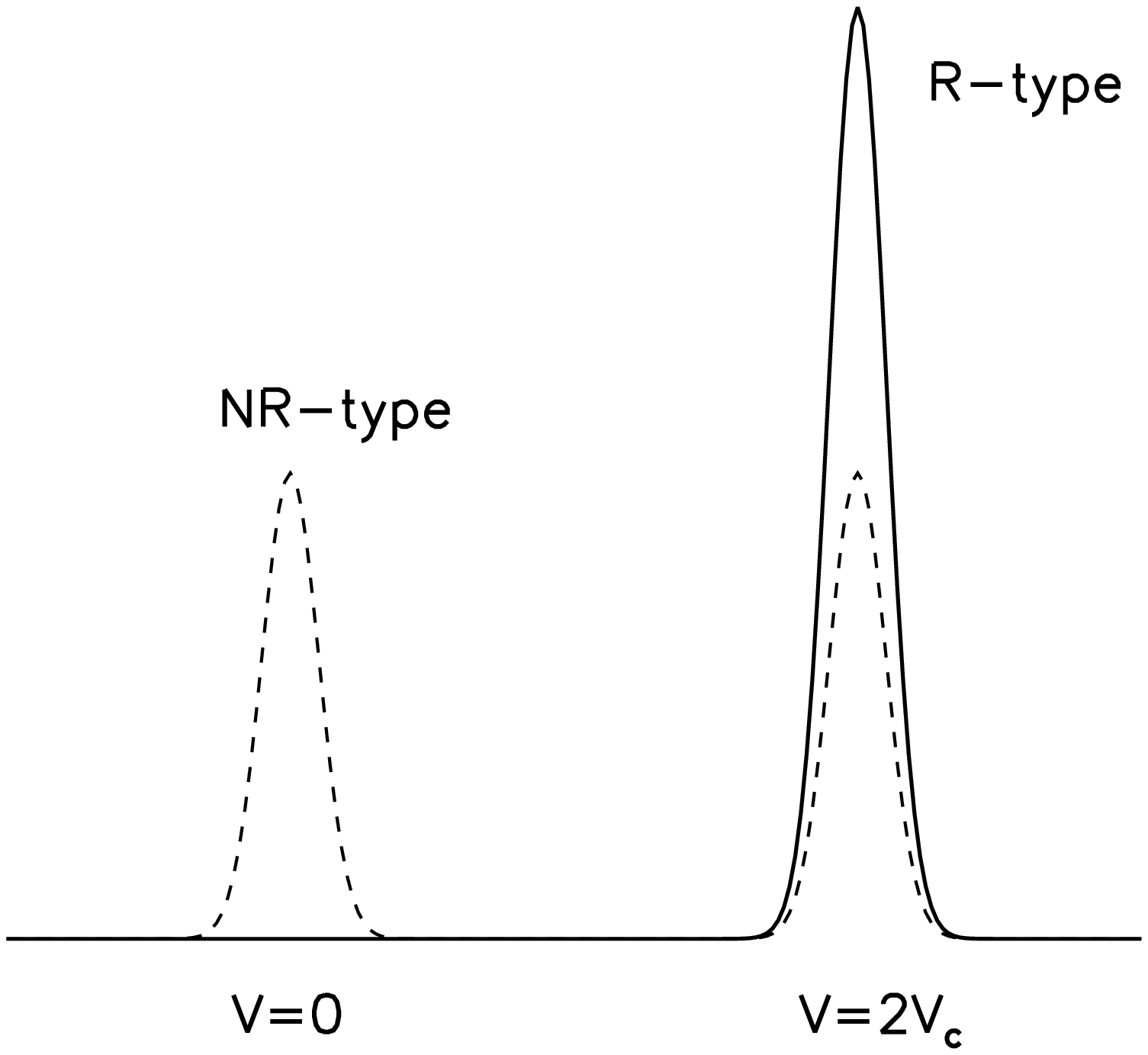}}
\caption{Two simple discs.  Models to the right (R-type model) are
rotating models 
where all stars (lenses and sources) move in the clockwise
direction, and models to the left are non-rotating with
half of the stars (lenses and
sources) orbiting in the clockwise sense, half anti-clockwise.  The
transverse velocity of the lens and the source at the line of sight to
the center (dotted lines) is $\pm V_c$.  The lower panel draws
schematic distributions of the relative source-lens speed
$V=|v_s-v_l|$ in the simplest case where the lens and the source are
at the front or on the back side of an annulus.  }
\label{physics}
\end{figure}
This is easily seen from the basic equation of microlensing,
\beq
\label{basictimescale}
{1 \over t} = {V \over R_E} = V \left({4 G m D\over c^2}\right)^{-1/2},
\qquad
D={D_l \over D_s} (D_s-D_l),
\eeq
where $G$ is the gravitational constant, $c$ is the velocity of light,
$R_E$ is the Einstein ring radius, $t$ is the event duration (one
Einstein radius crossing time), $D_l$ and $D_s$ are the distances to
the lens and the source, $V$ is the relative transverse speed between
the two, and $m$ is the mass of the lens.  For most microlensing
events one observes only their duration, $t$, from which one can
obtain a lensing probability (optical depth) $\tau$ and an event rate
$\Gamma$ observationally.  Other system parameters, such as the speed
$V$ and distances $D_s$ and $D_l$, can at best be inferred
statistically from a phase-space density model of the Galaxy.  As a
result the rate is related to the dynamical model as follows 
(cf. eq. [7] of Paczy\'nski 1986)
\beq
\label{averagerate}
{\Gamma \over \tau} = {2 \over \pi } \left< 1 \over t \right> 
                    = {c \over \pi \sqrt{G m D}} \left< V \right>,
\eeq
where the brackets indicate averages. In order to emphasize the effect
of the velocity distribution we have fixed the mass of the lenses $m$
and the positions of the lens and source.

Eq.~(\ref{averagerate}) can also be interpreted as a constraint on
lens mass.  If the event rate is fixed at the observed value and the
lens volume density model is fixed to fit the observed optical depth,
then eq.~(\ref{averagerate}) implies that
\beq
\label{mv2}
m \propto \left< V \right>^2, 
\eeq
so that the inferred lens mass is sensitive to the velocity
distribution of the dynamical model. An uncertainty of $1.2-2$ in
velocity translates to a factor of $1.4-4$ in the lens mass, in which
case one could mistake, e.g., an M-dwarf dominated model for a
brown-dwarf dominated model.

To illustrate this point, it is best to take one step backwards to
consider the simplest axisymmetrical models.  If one observes the two
discs shown in Figure \ref{physics} edge-on, then the distribution of
transverse relative source-lens speed peaks at $V=2V_c$ for the
maximum rotating disc (R-type; R stands for rotator), but peaks at
both $V=2V_c$ and $V=0$ for the two-stream counter rotating disc
(NR-type, standing for non-rotator).  Since the median speed $\left< V
\right>$ is twice as large in the R-type model as in the NR-type
model, twice as many events with duration twice as short are expected
in the R-type model as in the NR-type model.  In a more realistic
situation where lenses and sources are distributed across a uniform
disc and they orbit the center at constant angular speed, the R-type
model would predict (after a simple calculation) an average speed
$\left< V \right>$ and an event rate $\Gamma$ a factor of 1.2 times
longer than in NR-type models.

Past attempts to statistically determine the mass function of
microlenses all suffer from uncertainties in the dynamical models of
the inner Galaxy. Binney, Gerhard \& Spergel (1997) and Zhao (1997)
studied volume density models of the Galactic bar consistent with the
projected light distribution from the COBE/DIRBE maps, Zhao \& Mao
(1996, hereafter ZM) addressed the effects of the uncertain volume
density model on the microlensing probability (the optical depth).
This paper extends such studies to the microlensing event rate
distribution.  We concentrate on breaking the degeneracy of dynamical
models with the same three-dimensional density distribution but
different velocity fields, which are indistinguishable in studies
based on the projected light and the optical depth.

The main goal of this paper is to study the trend of the microlensing
event rate as a function of the key parameters of the bar (the pattern
speed, the axis ratio and the orientation angle) assuming a fixed lens
mass $m$.  This helps us to gauge whether the lens mass should
increase or decrease in case a modification of the bar parameters is
required to be consistent with some future observations or dynamical
models.  A side result is to give an error bar for predictions of the
lens mass due to the non-uniqueness of the bar.

The rigorous way to survey microlensing properties of Galactic bar
models is to build many sequences of three-dimensional bar models such
as in Zhao (1996) which cover the multi-parameter space. But such
heavy numerical modelling would be very inefficient. For a first study
it is more interesting to gauge the underlying physical effects with
simple theoretical models.  In this paper we show which insights can
be gained from studying the microlensing properties of the analytical
two-dimensional Freeman (1966) bars, which surprisingly capture most
of the microlensing effects of a realistic bar fairly well.

The outline of the paper is as follows.  In \S 2, we summarize the
properties of the Freeman bars, and we derive the optical depth for
self-lensing. In \S 3, we derive basic scaling relations for the event
rate distribution.  In \S 4 we show the run of event rate as functions
of bar parameters.  In \S 5 we show how to distinguish Freeman bars
with the same projected density and optical depth map, and in \S 6 we
derive the rate for extremely short events. Finally, we discuss
implications for the Galactic bar in \S 7. Some mathematical details
are given in the Appendix.

\section{Self-lensing of the Freeman bars}

Freeman (1966, hereafter F66) discovered self-consistent tumbling bar
models with a known analytical distribution function.  These
two-dimensional bars have been widely used to gain insight into the
structure of general self-consistent bars (Hunter 1974; Tremaine 1976;
Weinberg \& Tremaine 1983).  Despite the two-dimensional nature and
the special distribution functions, this class of easy-to-build bars
contains the main factors by which bars can affect the microlensing
rate.  These include its elongated density contours and similarly
elongated velocity ellipsoid, plus the pattern rotation and streaming
motion.  Our aim is to vary these factors and ``observe'' the bar both
at a variable line-of-sight angle from its long axis and at a variable
projected radius.

\subsection{Freeman bars}

The Freeman bar is a two-dimensional inhomogeneous elliptical bar of
finite extent, with surface density $\Sigma$ given by
\beq
\label{faceon}
\Sigma(X,Y) = {3 M \over 2 \pi a b} 
 \left(1-\phi\right)^{1 \over 2},
\qquad 
\phi={X^2 \over a^2}+{Y^2 \over b^2},
\eeq
and a potential
\beq
\label{potential}
\Phi(X,Y)={ G M \over 2 (a b)^{3 \over 2} } 
          [A_2(q) X^2+  B_2(q) Y^2], \qquad q \equiv {b \over a},
\eeq
where $M$ is the total mass, $a$ and $b$ are the semi-major ($X$) 
and semi-minor ($Y$) axes of the bar, and $A_2(q)$, $B_2(q)$ are
dimensionless function of the axis ratio $q$ given in equation (2) of
F66. The symmetry axes ($X$ and $Y$) rotate with an angular speed
$\Omega$ with respect to the rest frame.  For a given density, there
is a sequence of self-consistent models with different amounts of
pattern rotation. The sequence is a function of $\Omega/\Omega_J$,
where
\beq
\Omega_J^2(M,a,b) \equiv {G M \over (a b)^{3 \over 2} } 
                  { q^{-1} A_2- q B_2 \over  q^{-1} - q},
\eeq
is so defined that when $\Omega=\Omega_J$ the bar is of Jacobi type,
that is, with neither internal streaming motion nor velocity
anisotropy.

\begin{figure} 
\epsfysize=10cm 
\centerline{\epsfbox{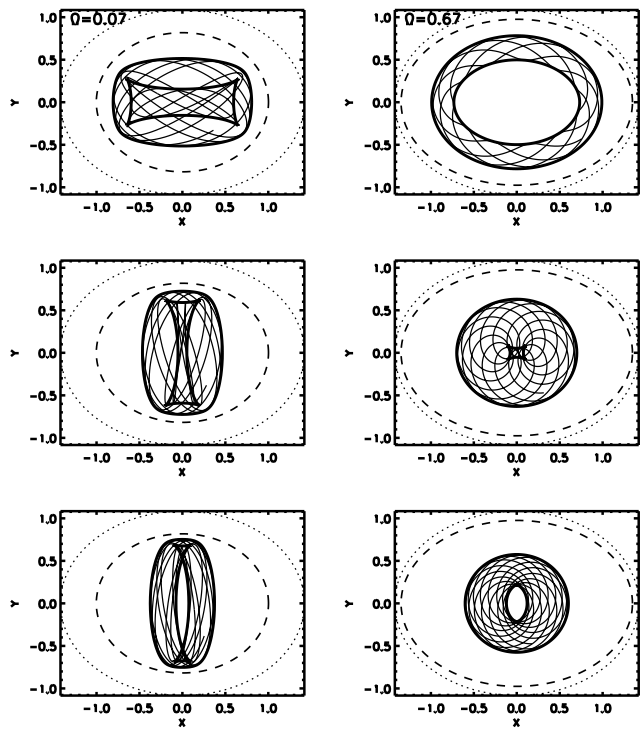}}
\caption{
Orbits in Freeman bars with $GM=1$ and semi-axes $a=1.420$ and $b=1.084$ 
and a boundary as indicated with the dotted ellipses in each panel.
The orbits are drawn in the $X-Y$ frame corotating with the tumbling bar 
with a pattern rotation speed
$\Omega=0.07$ (left panels) and $0.67$ (right panels).  
All orbits are assigned the same Jacobi integral $E_J=0.5$, which confines the
orbits within the zero velocity curves shown as the dashed ellipses in
each panel.  The heavy solid lines indicate the boundaries of each
orbit, as given in Appendix A. }
\label{orbits}
\end{figure}

The potential (\ref{potential}) is that of a two-dimensional
anisotropic harmonic oscillator with generally incommensurable
frequencies in the $X$ and $Y$ directions. The motion is separable in
canonical coordinates $(P_1, P_2, Q_1, Q_2)$ defined by F66.
Accordingly, all orbits have two independent isolating integrals of
motion, and they are in fact rectangular Lissajous figures in the
$(Q_1, Q_2)$-plane. In the corotating $(X, Y)$-frame, the orbits show
a more interesting variety of shapes, similar to what is seen in more
general bar potentials. We illustrate these in Figure \ref{orbits}.
The boundary curves of each orbit can be worked out analytically. We
give the equation in Appendix A.  For the fast tumbling bar, the
orbits range from direct to retrograde.  As the tumbling frequency
approaches zero, all orbits become rectangular boxes.  By properly
weighting orbits with different size and axis ratio, F66 was able to
produce a self-consistent elliptical bar.  Its distribution function,
which is positive definite, has the simple form
\bey
f(J) &=& f_0 (1-J)^{-{1 \over 2}}, 
         \qquad\mbox{\RM {if \, $0\le J \le 1$,}} \\ \nonumber
     & = & 0\qquad\qquad\mbox{\RM otherwise.}
\eey
The analytical expression for $J$ is fairly lengthy, but is
essentially a linear combination of the two energy integrals, and is
given in equations (43)--(45) of F66.

The system has a very simple velocity distribution.  In the frame
rotating with the bar, stars stream in the retrograde sense with
elliptical streamlines concentric to the bar's boundary.  The velocity
ellipsoid is generally anisotropic with the major axis either in the
$X$ or $Y$ direction.  The amplitude of the streaming motion is linear
with respect to the coordinates, and the amplitude of the dispersions
decreases to zero with a radial dependence $\sigma \propto
(1-\phi)^{1/2}$, identical to that of the surface density given in
eq.\ (\ref{faceon}).  A selfconsistent Freeman bar model is fully
specified by its mass $M$, the semi-axes $a$, $b$ and the pattern
speed $\Omega$.  One can construct a sequence of bar models from
needle-shaped to axisymmetric by varying the axis ratio $b/ a$. By
varying only $\Omega/\Omega_J$ one can build a sequence of models with
the same density distribution but different velocity
anisotropy and streaming motions.

\subsection{Distributions of the transverse velocity and the
            density in the line of sight}

The microlensing rate of the bar is determined by the distributions of
the transverse velocity and the three-dimensional volume density in
the line of sight path to a source.  We assume that the observer is
sufficiently far away from the center of the Freeman bar that all
perspective effects can be ignored.  Even though this makes the
analytical studies here not rigorously applicable to the Galaxy, where
perspective effects are a main clue to the size and orientation of the
bar, it still provides very useful insights.

\begin{figure} 
\epsfysize=8cm \centerline{\epsfbox{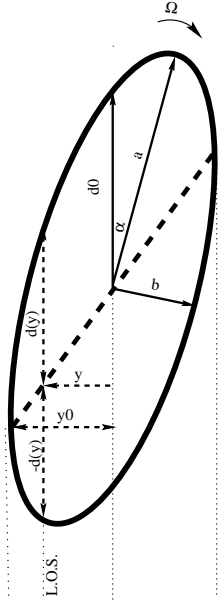}}
\caption{
The different coordinate systems used.  $\tilde{x}=0$
along the heavy dashed line.} 
\label{coordinates}
\end{figure}

The density distribution depends on the thickness of the bar, on the
orientation of the bar and on the impact parameter of the line of
sight of the observer. To modify the 2D Freeman bar to a
three-dimensional model, we ``puff up'' the Freeman bar by a small
constant thickness $\Delta$, so that the volume density distribution
is given by\footnote{The non-axisymmetric perturbation in the
COBE/DIRBE map of the Galaxy has a thickness no less than the thin
disc ($\Delta>2 \times 200$pc) and semi-minor axis no greater than the
corotation radius ($b<a<4$kpc) so the requirement that $\Delta/b \ll 1
$ is barely satisfied.}
\beq
\rho(X,Y,Z)={\Sigma(X,Y) \over \Delta},
\qquad
\Delta \ll {\rm min}(a,b).
\eeq
ZM use a Galacto-centric Cartesian coordinate system $(x,y,z)$ with
the $x$-axis pointing away from the Sun, making an angle $\alpha$ with
the bar's long ($X$) axis. Similar but more convenient is the
following coordinate system $(\tilde{x}, y, z)$ with
\beq
\tilde{x} \equiv x - C y, 
\qquad
C={(q-q^{-1}) \sin \alpha \cos \alpha 
              \over q \cos^2 \alpha +q^{-1} \sin^2 \alpha },
\eeq
so that it is obtained by applying a shear to the original $(x,y,z)$
coordinate in the line-of-sight direction, as illustrated in Figure 
\ref{coordinates}. In these coordinates the face-on density of the 
Freeman bar can be rewritten as
\beq
\label{surf}
\Sigma(\tilde{x},y) = {3 M \over 2 \pi a b} 
\left(1- { \tilde{x}^2 \over d_0^2} - {y^2 \over y_0^2} \right)^{1 \over 2},
\eeq
where $2 y_0$ and $2 d_0$ are the projected width and the central
depth of the bar. Both quantities are functions of $a$, $b$ and
$\alpha$, given by
\beq
y_0= \left(a b \gamma \right)^{1 \over 2},
\qquad
d_0= \left({a b \over \gamma} \right)^{1 \over 2},
\eeq
with 
\beq
\label{defgamma}
\gamma (\alpha, q) \equiv q^{-1} \sin^2 \alpha +q \cos^2 \alpha.
\eeq
A line of sight with an impact parameter $y$ intersects with the bar's
near end at $\tilde{x}=-d(y)$ and far end at $\tilde{x}=+d(y)$, where
\beq
d(y)= d_0 \left(1-{y^2 \over y_0^2} \right)^{1 \over 2},
\eeq
so that $2d(y)$ is the depth of the bar at $y$.  For an axisymmetric
disc, $a=b=y_0=d_0=\sqrt{d(y)^2+y^2}$, and $\gamma=q=1$.

The transverse velocity distribution depends on the velocity field of
the bar, the orientation and the impact parameter of the line of
sight. It is given by a box car distribution: 
\bey\label{pv}
P(v_y) &=&\int f(J) \d v_x \\ \nonumber 
&=& {1 \over 2 w } \qquad \mbox{\RM{if $|v_y-\bar{v}| \le w$,}}\\  \nonumber
&=& 0 \quad\qquad \mbox{\RM{otherwise,}}
\eey
where the local transverse streaming velocity $\bar{v}$ is a linear
function of $(x,y)$, and the half width $w$ is $\sqrt{3}$ times the
local transverse dispersion, which is proportional to $(1-\phi)^{1
\over 2}$.  $\bar{v}$ can be split into a component $\tilde{v}$
proportional to $\tilde{x}$ and a (less useful) component proportional
to $y$.  More rigorously
\beq\label{wvbarscale}
{\tilde{v} \over \sigma(y)} = \lambda {\tilde{x} \over d(y)}, \qquad
{w \over \sigma(y)} = \sqrt{3} \left( 1 - {\tilde{x}^2 \over d^2(y)} 
                                                   \right)^{1 \over 2},
\eeq
where we have rescaled the $\tilde{x}$ coordinate by $d(y)$, the $y$
coordinate by $y_0$ and the velocities by $\sigma(y)$, which is
defined as
\beq
\label{sigmay}
\sigma(y) \equiv \sigma_0 \left(1-{y^2 \over y_0^2} \right)^{1 \over 2},
\eeq
and $\sigma_0=\sigma(y=0)$ is the central transverse dispersion.  The
parameter $\lambda$ is defined by
\beq\label{lambda}
\lambda\left({b \over a}, \alpha, {\Omega \over \Omega_J}\right)
 \equiv {\bar{v}_{\rm max} \over \sigma_0},
\eeq
where $\bar{v}_{\rm max}$ is the transverse rotation speed at $\tilde{x} =
\pm d_0$ along the $y=0$ line.  So $\lambda$ is a dimensionless global
(independent of $y$) indicator of the amount of rotational support in
the system (similar to the conventional $V_{\rm rot}/\sigma$ parameter, 
e.g., Binney \& Tremaine 1987). Both $\sigma_0$ and $\lambda$ are lengthy
functions of the bar axis ratio and pattern speed, which are given in
Freeman (1966) in the case that the bar angle $\alpha=0^o$ or $90^o$.
The dependence on $\alpha$ is then easily worked out with
\beq
\label{angledependence}
\left[ \begin{array}{c} 
\! \sigma_0^2(\alpha) \! \\ 
  (\lambda \sigma_0)(\alpha) \!
\end{array} \right]
=
\left[ \begin{array}{cc} 
\! \sigma^2_0(0) & \sigma^2_0(90) \! \\
\! (\lambda \sigma_0)(0) & (\lambda \sigma_0) (90) \!
\end{array} \right]
\left( \begin{array}{c} 
\!\cos^2 \alpha \! \\ \!\sin^2 \alpha \!
\end{array} \right).
\eeq
For our purpose it is only important to know that {\sl for a
non-rotating bar $\lambda=0$ and the velocity ellipsoid is aligned
with the bar; for a Jacobi-type rotator ($\Omega=\Omega_J$), stars have
an isotropic dispersion on top of a solid body pattern rotation with
respect to the rest frame.}

More relevant to the microlensing is the lens-source relative
transverse speed $V$ and its distribution $F(V)$, which is a
convolution of the distribution $P(v_y)$ of eq.~(\ref{pv}) for the
transverse velocity distribution of the lens or the source.
\beq
\label{fvdef}
F(V) = \int P(v_y) \left[ P(v_y-V) +P(v_y+V)\right] \d v_y,
\eeq
where we have integrated over the source velocity $v_y$.  The
integration can be carried out, but the result is somewhat lengthy,
and is given in eq.~(\ref{FV}).  We remark that $F(V)$ satisfies the
normalization $\int_0^{+\infty} F(V) \d V =1$.

\subsection{Optical depth}

Following ZM, we assume that the sources are distributed with a number
density proportional to the mass density $\rho_s$, which is in turn
proportional to $\Sigma_s$. Then the lensing optical depth averaged
over all sources along the line of sight with impact parameter $y$ is
\beq
\label{tauf}
\tau(y) =  \int\limits_{-d(y)}^{+d(y)} \! \d\tilde{x}_s \, \Sigma_s \!\!\!\!
           \int\limits_{-d(y)<\tilde{x}_l<\tilde{x}_s}  \!\!\! \d\tau_s \,\,
          \Big/  \int\limits_{-d(y)}^{+d(y)} \! \! \d\tilde{x}_s \, \Sigma_s
\eeq
where
\beq
\label{taus}
d\tau_s= {4\pi G \over c^2} 
(\tilde{x}_s-\tilde{x}_l) \rho_l d\tilde{x}_l,
\eeq
is the optical depth for a source located at $(\tilde{x}_s,y,0)$, and
$\rho_l$ is the volume density at the lens position.  Here the
subscripts $_s$ and $_l$ denote quantities of the source and the lens,
respectively. For the Freeman bar:
\beq
\tau(y) = {128 G \over 15 \pi c^2} {M \over \Delta} {1 \over
\gamma }
\left(1 - {y^2 \over y_0^2} \right)^{3/2}. 
\eeq

\section{Event rate distribution}

\subsection{Definitions}

The duration of a microlensing event is related to the lens mass $m$
and to the relative distance $\tilde{x}_s-\tilde{x}_l$ and velocity
$V$ between the lens and the source by (cf.\ eq.~\ref{basictimescale})
\beq
t \approx {1 \over V} \sqrt{{4 G m \over c^2} 
\left( \tilde{x}_s-\tilde{x}_l \right) },
\eeq
where we have made the approximation $D \approx
\tilde{x}_s-\tilde{x}_l$ for self-lensing of a far-away bar.  If one
fixes the lens mass $m$ and the source position $\tilde{x}_s$, then
the probability of observing an event with any duration is the optical
depth $d\tau_s$ (cf.\ eq.\ \ref{taus}).  But only the fraction
$F(V)dV$ of the lenses with relative velocity $V$ to $V+dV$
contributes to the event with duration $t$ to $t+dt$, where $F(V)$ is
the relative velocity distribution given in eq.\ (\ref{fvdef}).  So
the contribution $d\Gamma$ to the event rate is
\beq
d\Gamma = {2 \over \pi t} \, d\tau_s F(V)dV,
\eeq
where $2 /(\pi t)$ is the average frequency of an event with time
scale $t$, and $2/\pi$ is the ratio of the Einstein diameter to the
area in dimensionless units (Paczy\'nski 1991). It follows that the
differential duration distribution is given by
\beq
{d\Gamma \over d\log t} = {2  \ln 10 \over \pi t} d\tau_s F(V)V.
\eeq
Just as for the case of the optical depth (\ref{tauf}), the observable
rate should be averaged over the source distribution along the line of
sight. We define the microlensing duration distribution profile
$f(\log t)$ normalized by the optical depth as
\beq 
\label{flogt}
f(\log t) \equiv  {d\Gamma \over \tau d\log t}
\eeq
so that
\beq
\label{norm}
\int f(\log t) \d \log t  =  {\Gamma \over \tau},
\eeq
where $\Gamma$ is the total event rate. Then the source density
averaged duration profile is given by (cf.\ eqs \ref{tauf} and
\ref{taus})
\beq
f(\log t) = {2 \ln 10 \over \pi t} 
{\int\limits_{-d(y)}^{+d(y)} \! \d \tilde{x}_s \!\!
 \int\limits_{-d(y)}^{\tilde{x}_s} \! \d \tilde{x}_l 
                       (\tilde{x}_s\!-\!\tilde{x}_l) \Sigma_s \Sigma_l V F(V)
\over \int\limits_{-d(y)}^{+d(y)} \d \tilde{x}_s 
      \int\limits_{-d(y)}^{\tilde{x}_s} 
              \d \tilde{x}_l (\tilde{x}_s\!-\!\tilde{x}_l) \Sigma_s \Sigma_l }.
\eeq
Note that $f(\log t)$ is independent of the bar thickness $\Delta$.

\subsection{Scaling relations}

\begin{figure} 
\epsfysize=6cm \centerline{\epsfbox{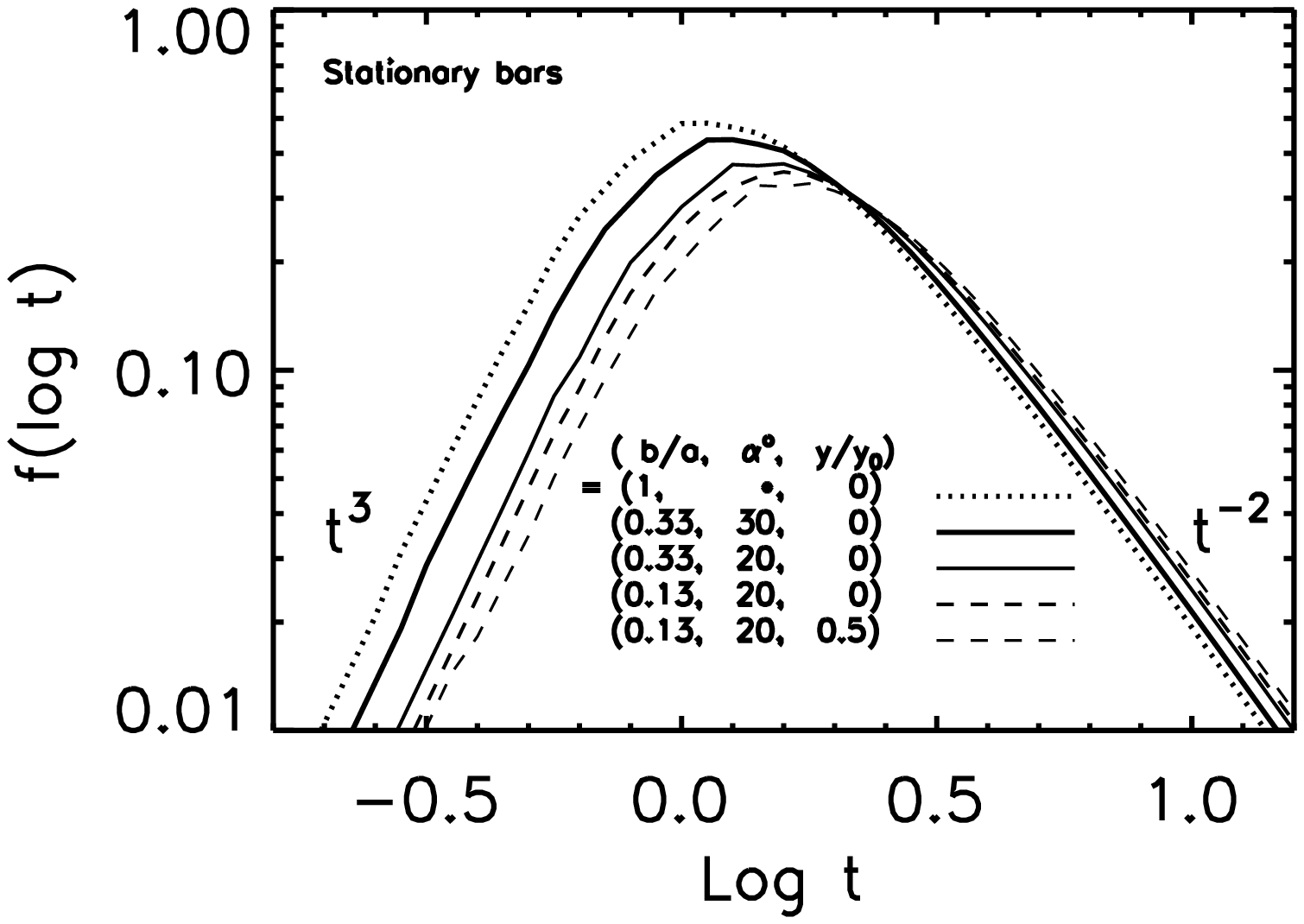}}
\epsfysize=6cm \centerline{\epsfbox{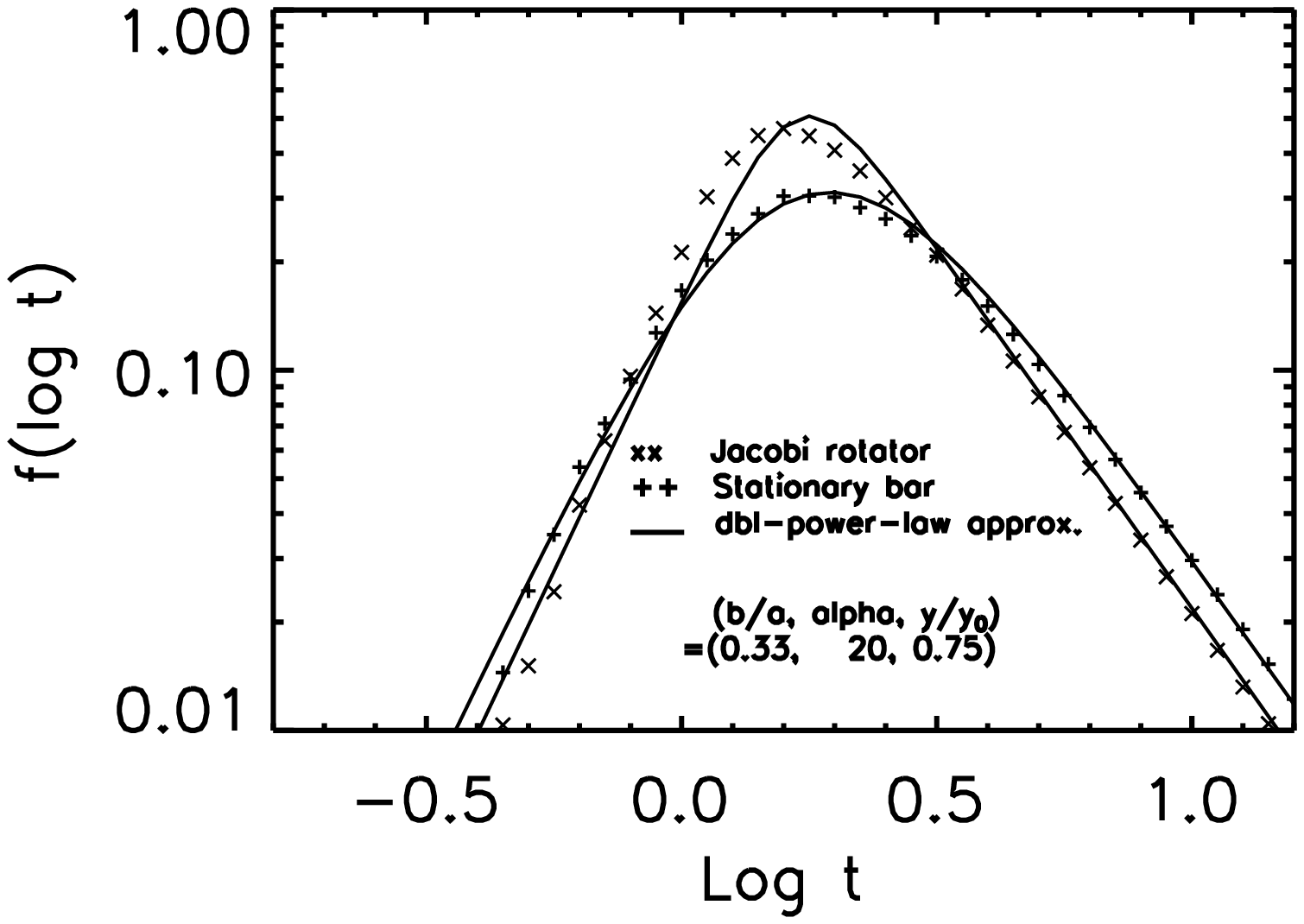}}
\caption{Upper panel: the event duration distribution for 
stationary Freeman bars 
with various axis ratio, angle, and impact parameter 
$(b/a, \alpha, y/y_0)$; the
curves are labeled in the sequence of their peak positions 
(from top left to bottom right).  The curves all have
the same shape, in particular, the same
asymptotic power-laws for very short $f(\log t) \propto t^3$
or very long events $f(\log t) \propto t^{-2}$.
Lower panel: a
Jacobi-type rotating model ($\Omega=\Omega_J$) compared with a
non-rotating one ($\Omega=0$), and the simple double-power-law
approximation to the distribution (cf.\ eq.~\protect{\ref{fit}}).  The
rotating models have a narrower distribution and higher rate.
}\label{profile}
\end{figure}

The normalized event duration distribution $f(\log t)$ should have the
dimension of a to-be-defined typical frequency $\nu$, multiplied by a
function of the dimensionless time $\nu t$.  We write this concisely
as
\beq\label{scaling}
f(\log t) = \nu g_{\lambda} (\nu t),
\eeq
where
\beq
\nu \equiv {\sigma(y) \over [{G m \over c^2} d(y) ]^{1 \over 2}} 
      = {\sigma_0 \over \left({G m \over c^2} d_0 \right)^{1 \over 2} }
   \left( 1-{y^2 \over y_0^2} \right)^{1 \over 4},
\eeq
and $1/\nu$ is a typical time scale for the events, $\lambda$ is the
ratio of rotation vs.\ random motion in the model, and
$g_{\lambda}(\mu)$ is a dimensionless function of $\mu$ whose
functional form depends only on $\lambda$, which is a dimensionless
function of $b/a$, $\Omega/\Omega_J$, and $\alpha$.  The exact
derivation of how $f(\log t)$ depends on $\lambda$ is given in
Appendix B.

Eq.~(\ref{scaling}) can be understood intuitively as follows.
Consider a line of sight with impact parameter $y$ to a stationary
Freeman bar model.  The lens-source distance $(\tilde{x}_s -
\tilde{x}_l)$ scales with the bar's depth $d(y)$, and the transverse
velocity $V$ scales with the dispersion $\sigma(y)$ (cf.\ eq.\
\ref{wvbarscale}).  So the Einstein radius 
$R_E \propto \left[m d(y) \right]^{1/2}$ and the event rate scale with
${1 \over t} = {V \over R_E} \propto {\sigma(y) \over \left(m
d(y)\right)^{1/2}} \propto \nu $.  As a result, the event duration
distribution depends on the bar density parameters, angle and impact
parameter only through $\nu$.  For rotating bars, the profiles also
depend on the dimensionless quantity $\lambda$. 

A surprising but direct result of the scaling relation
eq.~(\ref{scaling}) is that all non-rotating Freeman bars have the
same ``shape'', as illustrated in Fig.~\ref{profile}.  This means that
although the $\log t$ vs. $\log f$ diagram of models with the same
$\lambda$ have different median event duration and total rate, they
will coincide after shifting the zero points in both axes by a
constant $\log \nu$.  In particular, there is a one-to-one
correspondence between a non-rotating Freeman bar with a non-rotating
Kalnajs (1976) disc, which is an axisymmetric version of the Freeman
bar.

Interestingly the event duration distributions of all Freeman bars
with the same $\lambda$ also should have the same ``shape''.  The
shape is a function of $\lambda$ only.  For non-rotating models
$\lambda=\Omega=0$.  One can always find a Kalnajs disc with the same
rotation-to-dispersion ratio ($\lambda$) and the same shape of the
microlensing event duration distribution.  Just as for the
dimensionless $\lambda\left({b \over a}, \alpha, {\Omega \over
\Omega_J}\right)$, the width of the event duration distribution
depends on the bar angle and axis ratio directly, but depends on the
size and mass of the bar only through the normalized pattern speed
$\Omega/\Omega_J(M,a,b)$.

The shape of the event duration profiles is also invariant when the
same bar model is viewed at different impact parameters because the
global parameter $\lambda$ is independent of the impact parameter $y$
(cf.\ Fig.~\ref{profile}).

\subsection{Asymptotic behaviour}

Now we consider the asymptotic behavior of the event time scale
distribution.  The asymptotic power-law profiles are evident in
Figure~\ref{profile}.  This is because {\sl very short events come
from a pair of lens and source which are very close in distance, and
very long events happen if they are very close in proper motion.}  Mao
\& Paczy\'nski (1996) show that this results in two generic power-law
profiles for $f(\log t)$ at very small or large values of $t$ for
three-dimensional models.  Similar to these authors, we found the
following asymptotic relations for the two-dimensional Freeman bar,
\footnote{for very long duration, $f(\log t)$ generally is proportional 
to $(\nu t)^{-dim}$, where $dim$ is the dimension of the system.  So
for three-dimensional bars, ${1 \over \nu} f(\log t) \propto (\nu
t)^{-3} $ when $t \rightarrow \infty$.}
\bey
{1 \over \nu} \cdot f(\log t)  &=& g_{\lambda}(\nu t)\\ \label{shortside}
&\rightarrow & {108 \ln 10 \over 35 \pi}\left( \nu t \right)^3 
 \qquad \mbox{$t\rightarrow 0$},\\ \label{longside}
&\rightarrow & {108 \ln 10 \over 35 \pi } \xi(\lambda)^5 
        \left( \nu t \right)^{-2} \qquad 
  \mbox{$t\rightarrow \infty$},
\eey
where $\xi$ is a dimensionless function of $\lambda$ with
$\xi(0)=0.94$ (cf.\ eq.~[\ref{fitstuff}]).  

The full profile of $f(\log t)$ can be constructed, approximately, by
interpolating between the asymptotic relations 
\bey
{1 \over \nu} &\cdot& f(\log t) = g_{\lambda}(\nu t)\\ \label{fit}
                              & \approx & 
{108 \ln 10 \over 35 \pi}\left( \mu \right)^3 \left[
1+\left( \mu \over \xi \right)^{1 \over p} \right]^{-5 p}, 
\quad \mu \equiv \nu t,
\eey
where $p$ is a measure of the width of the distribution.  Both $p$ and
$\xi$ are weakly decreasing functions of $\lambda$.  We find that
\beq\label{fitstuff}
p(\lambda) \approx 0.36 \left(1-{\lambda^2 \over 2}\right),
\qquad 
\xi(\lambda) \approx 0.94 \left(1+{\lambda^2 \over 5}\right)^{-{1 \over 3}},
\eeq
together with eq.~(\ref{fit}), give a reasonably good approximation to
$f(\log t)$, good within 10\% for non-rotating Freeman bars
$(\Omega=\lambda=0)$ (cf.\ Fig.~\ref{profile}).  Using this
interpolation, we find that $\log t$ has a mean at approximately
$-0.14 - \log \nu$ dex with an rms width approximately $0.32$ dex;
both the mean and the width are also weak functions of $\lambda$.

The function $f(\log t)$ is independent of $\lambda$ or rotation for
very short events because the lens and source of these events are very
close in distance, so their relative rotation speed is always zero.
The short duration events are particularly useful for constraining the
mass of the lenses in the bar because lenses in the foreground disc
are generally sufficiently far away from the sources in the bar that
they do not contribute significantly to the short events.

Compared to non-rotating models, the $f(\log t)$ profiles are slightly
narrower for models with increasing rotation.  This can be understood
as a smaller spread in the velocities of lens and source in rotating
models leads to a smaller spread of the event duration.  However, the
variation of the width is small (see Fig~\ref{profile}), and to good
approximation Freeman bars all have very similar $f(\log t)$ profiles.

\section{The total event rate as function of the bar parameters}

\subsection{The size and mass of the bar}

The total event rate per optical depth is given by eqs.~(\ref{norm})
and~(\ref{scaling}):
\beq
\label{rate}
{\Gamma \over \tau}  = \int \nu g_{\lambda}(\nu t) \d (\log t)  
                     =  \xi_1 (\lambda) \nu,
\eeq 
where $\xi_1$ is a dimensionless increasing function of $\lambda$.  We find
\beq
\xi_1 (\lambda) \approx 0.4 \left(1+{\lambda^2 \over 5}\right).
\eeq
Eq.~(\ref{rate}) shows that ${\Gamma \over \tau}$ is a function of
$\nu$ and $\lambda$ only, and it is proportional to $\nu$.  We now
show that several useful relations follow from eqs.~(\ref{rate}) and
(\ref{scaling}).

The event rate scales with the lens mass $m$, the bar mass $M$, size
$L \equiv \sqrt{a b}$, and thickness $\Delta$.  We define the following
characteristic lens-source velocity $V^{*}$, Einstein radius
$R^{*}_E$, and frequency $\nu^{*}$,
\bey
\label{nuscale}
V^{*}   &\equiv&   \sqrt{G M \over L}, \\ \nonumber
R^{*}_E &\equiv& \sqrt{{G m \over c^2} L }, \\ \nonumber
\nu^{*} &\equiv& {V^{*} \over R^{*}_E } = {c \over L} \sqrt{M \over m}. 
\eey
Then
\beq
{\Gamma \over \tau} \propto \nu^{*} 
                = {c \over \sqrt{a b} } \sqrt{M \over m}.
\eeq
Upon substitution in eq.~ (\ref{tauf}), we find the following scaling 
relation for the optical depth and the event rate,
\beq\label{tau-gamma-scale}
\tau \propto \tau^{*} \equiv {G \over c^2} {M \over \Delta},
\qquad
\Gamma \propto \Gamma^{*} \equiv {G \over c} {M \over \sqrt{a b} \Delta} 
                                                 \sqrt{M \over m}.
\eeq
For a compact bar with high $M$ or low $a b$, the velocity dispersion
is high, so the event duration is shortened, and the event rate is
increased as expected.

\begin{figure} 
\epsfysize=6cm \centerline{\epsfbox{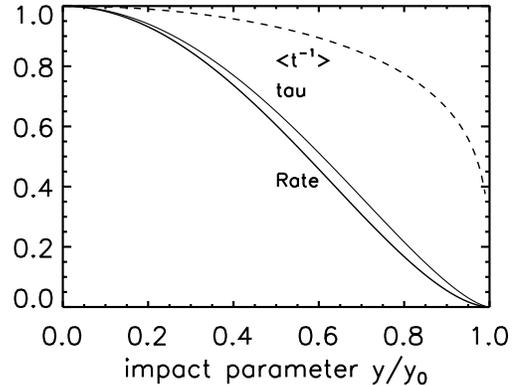}}
\caption{The optical depth, event rate, and event rate per optical depth
(scaled with values at $y=0$) as functions of the impact parameter
$y$.  The behavior is the same for all Freeman bars.}
\label{impact}
\end{figure}

\subsection{The impact parameter of the line of sight of the observer}

Moving away from the center within a model, the event rate drops
faster than the optical depth, mostly as a result of a lower escape
velocity at large radii, so that the mean event duration $\propto
\nu^{-1}$ shifts towards larger values (cf.\ eq.~\ref{scaling}, and
Fig.~\ref{impact}). Specifically:
\beq
{\tau \over \tau^{*} } \propto 
            \left( 1-{y^2 \over y_0^2} \right)^{3 \over 2},
\eeq
\beq
{\Gamma \over \Gamma^{*} } \propto 
            \left( 1-{y^2 \over y_0^2} \right)^{7 \over 4},
\eeq
\beq
{\Gamma \over \tau \nu^{*} } \propto 
            \left( 1-{y^2 \over y_0^2} \right)^{1 \over 4}.
\eeq
However, the shape of the duration distribution profile is always
independent of the impact parameter (cf.\ eq.~\ref{scaling} and
Fig~\ref{profile}).

\subsection{The bar axis ratio and orientation angle}

\begin{figure} 
\epsfysize=6cm \centerline{\epsfbox{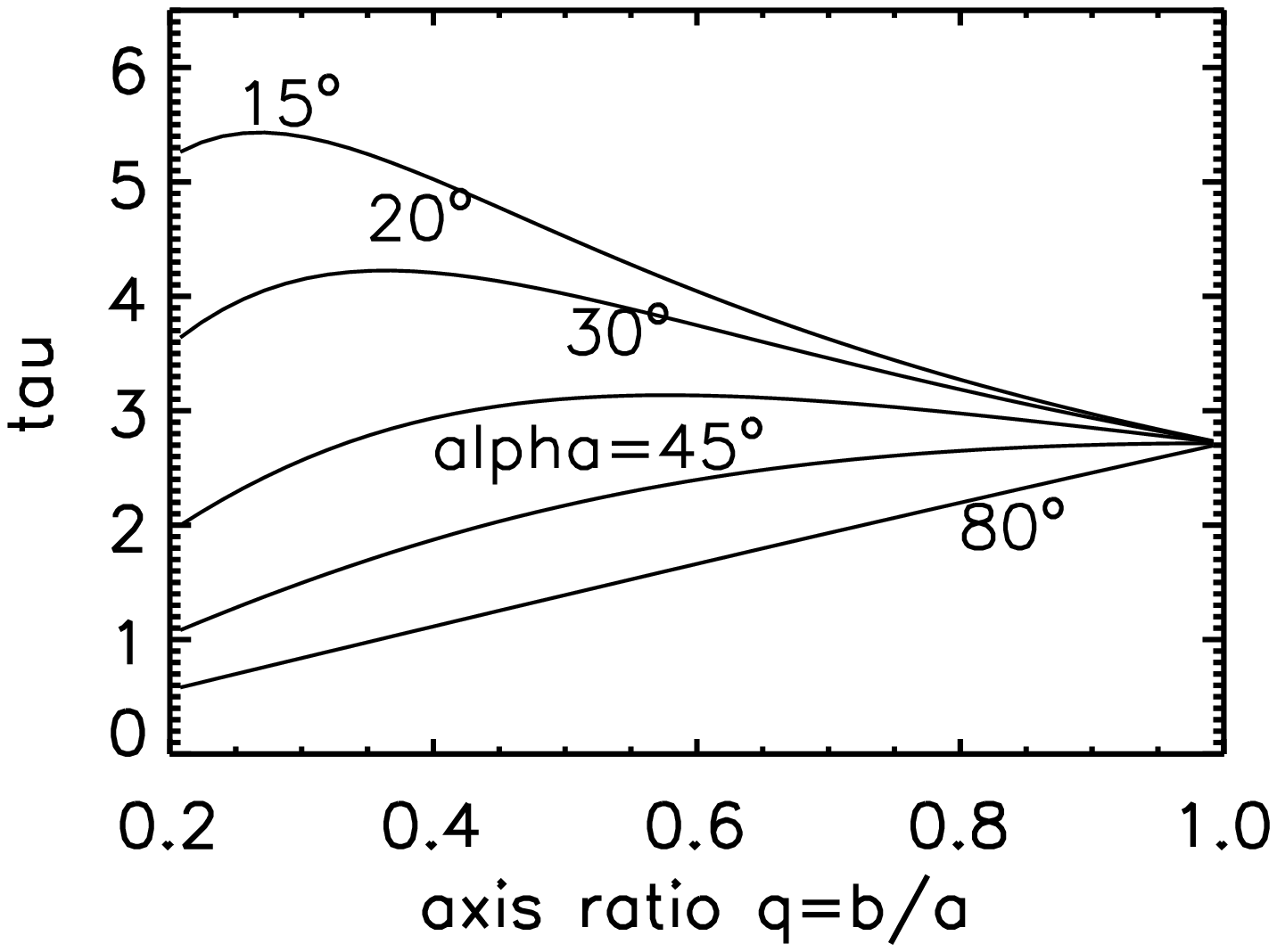}}
\epsfysize=6cm \centerline{\epsfbox{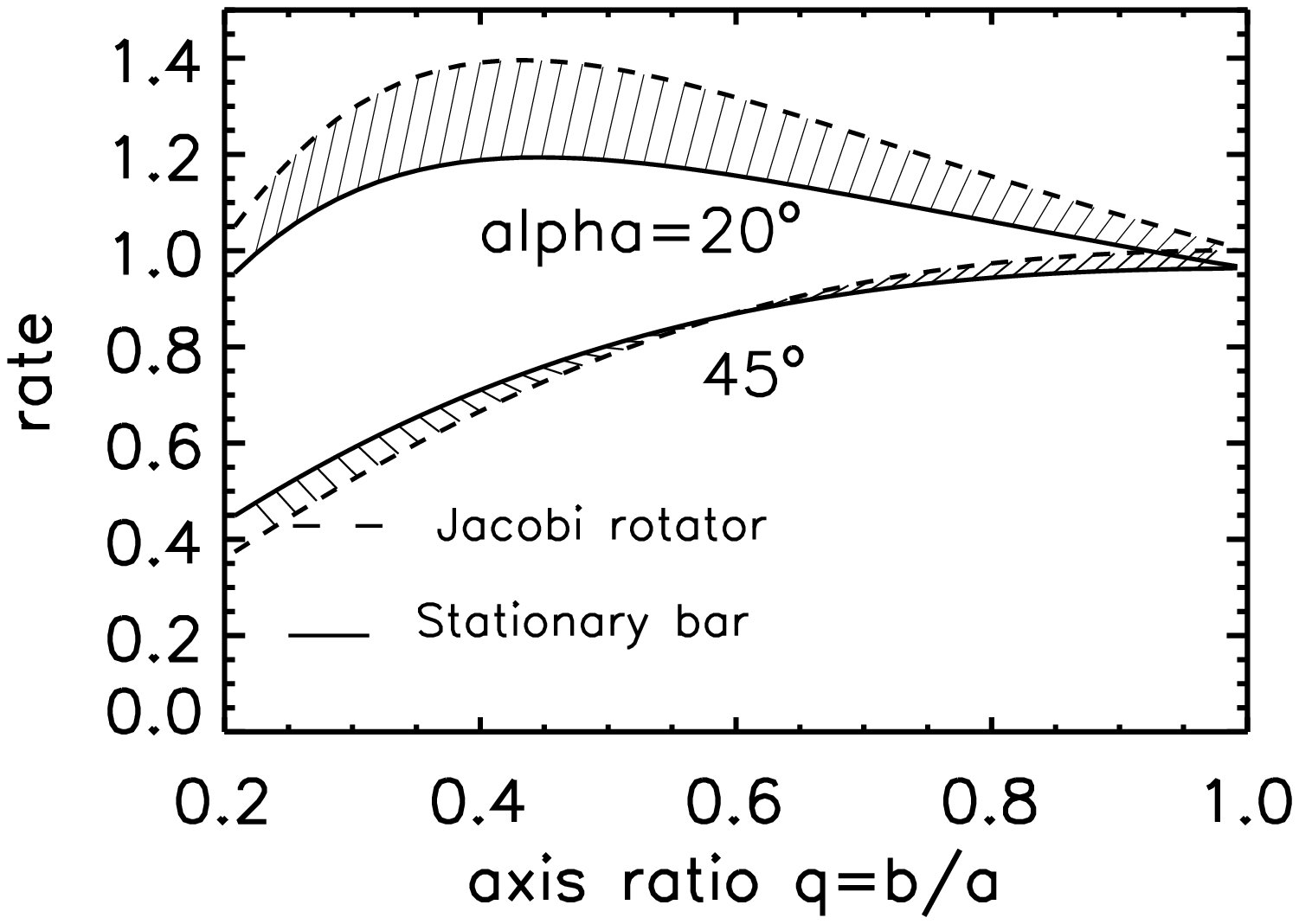}}
\epsfysize=6cm \centerline{\epsfbox{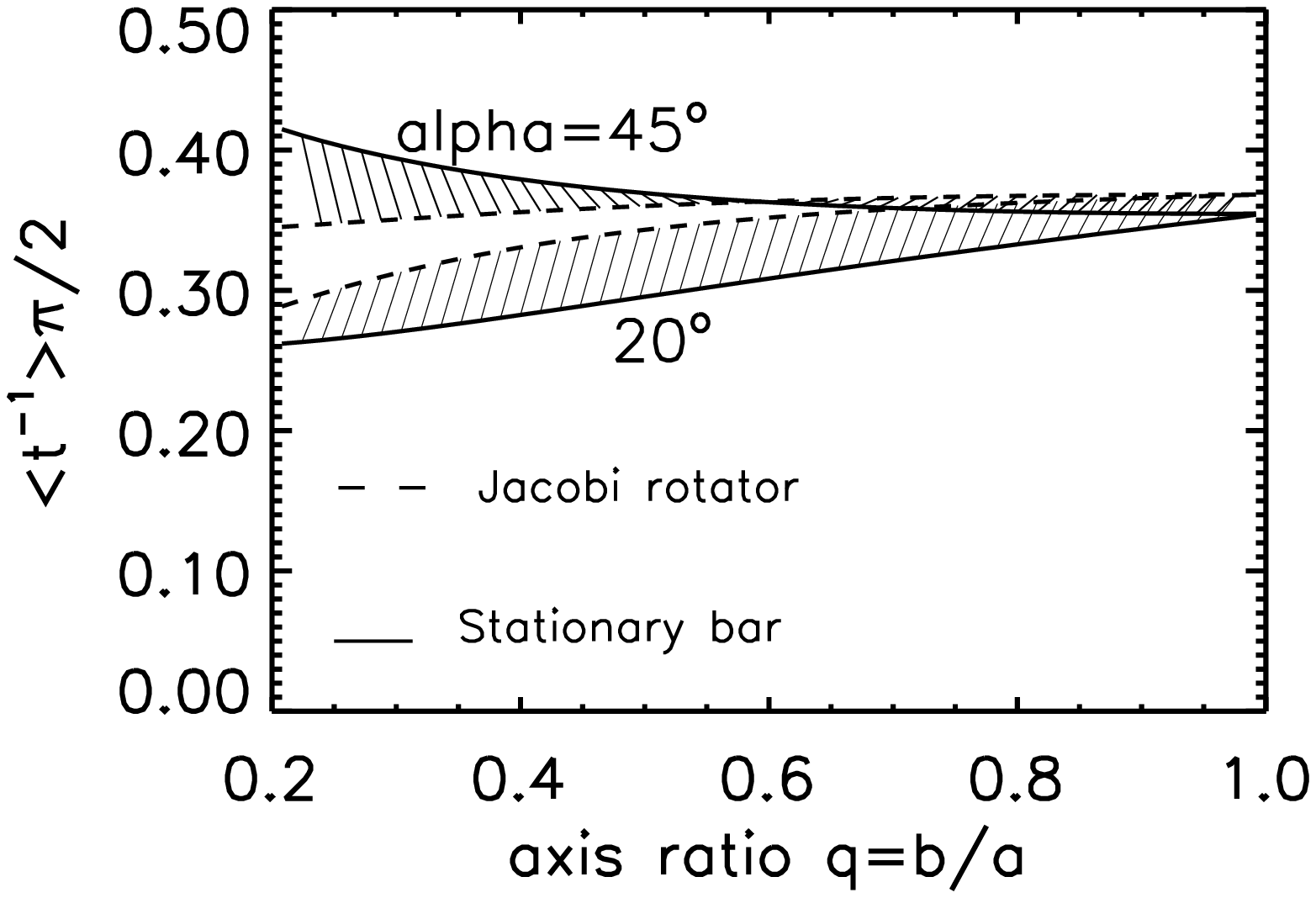}}
\caption{The optical depth (upper panel), event rate $\Gamma$ (middle 
panel) and event rate per optical depth $\Gamma \over \tau$ (lower
panel) as functions of the axis ratio of the bar at several values of
the bar angle for a line of sight through the center of the bars.  The
shaded areas show the spread due to different pattern speed.  $\Gamma$
and $\tau$ have been scaled with $\Gamma^{*}$ and $\tau^{*}$ (cf.\
eq.~\ref{tau-gamma-scale}). }
\label{angle-ratio}
\end{figure}

For the same bar axis ratio, and the same amount of rotational support
(say $\lambda=0$ as for stationary bars), the optical depth and the
event rate increase if the bar points closer to the line of sight, but
the events are longer (cf.\ Fig.~\ref{angle-ratio}).  This is because
when the bar points towards us, it is longest in the line of sight,
and smallest in transverse velocity (motions are primarily along the
long axis).  As a result, both the Einstein radius and the event
duration are at their maximum.  Figure~\ref{angle-ratio} also shows that
the event rate is significantly higher for a rotating model than for a
stationary bar with the same axis ratio and bar angle.

At a fixed angle less than $45^\circ$, there is an optimal axis ratio
(near $b/a = \tan \alpha$) for the bar to yield the highest rate and
optical depth (see also Fig.\ 2a of ZM, and Fig~\ref{angle-ratio} here).
At angles larger than $45^\circ$, the rate and the optical depth are
highest for the axisymmetric model.

Finally, the event rate $\Gamma$ largely follows the trend of $\tau$
as expected.  This is because
\beq
{\Gamma \over \tau} \propto \nu 
                    \propto \left({M \over a b}\right)^{1 \over 2} 
\gamma^{1 \over 4} \left( 1-{y^2 \over y_0^2} \right)^{1 \over 4},
\eeq
where $\gamma (\alpha, {b \over a})$ is given in eq.\
(\ref{defgamma}), and is generally a ``weaker'' function of the bar
parameters $(M, a, b)$ and the line-of-sight parameters $(\alpha, y)$
than is the optical depth
\beq
\tau \propto {M}
\gamma^{-1} \left( 1-{y^2 \over y_0^2} \right)^{3 \over 2}.
\eeq

\section{Total event rate for models with the same optical depth 
         and projected density}

Now we return to the question raised in the Introduction: how does the
event rate vary for models with the same optical depth and projected
density?  This is interesting because the projected density from COBE
maps and the observed optical depth of a bar do not uniquely specify
the volume density of the bar.  Also the velocity structure of the bar
is not uniquely constrained by self-consistency.  If one tries to
derive the lens mass by fitting the observed microlensing event rate
with any specific model, the result will be model-dependent.  Here we
try to study this systematic effect.

\subsection{Pattern speed}

First, one can generate a sequence of Freeman bars with the same
density but different pattern speeds.

\begin{figure} 
\epsfysize=6cm 
\centerline{\epsfbox{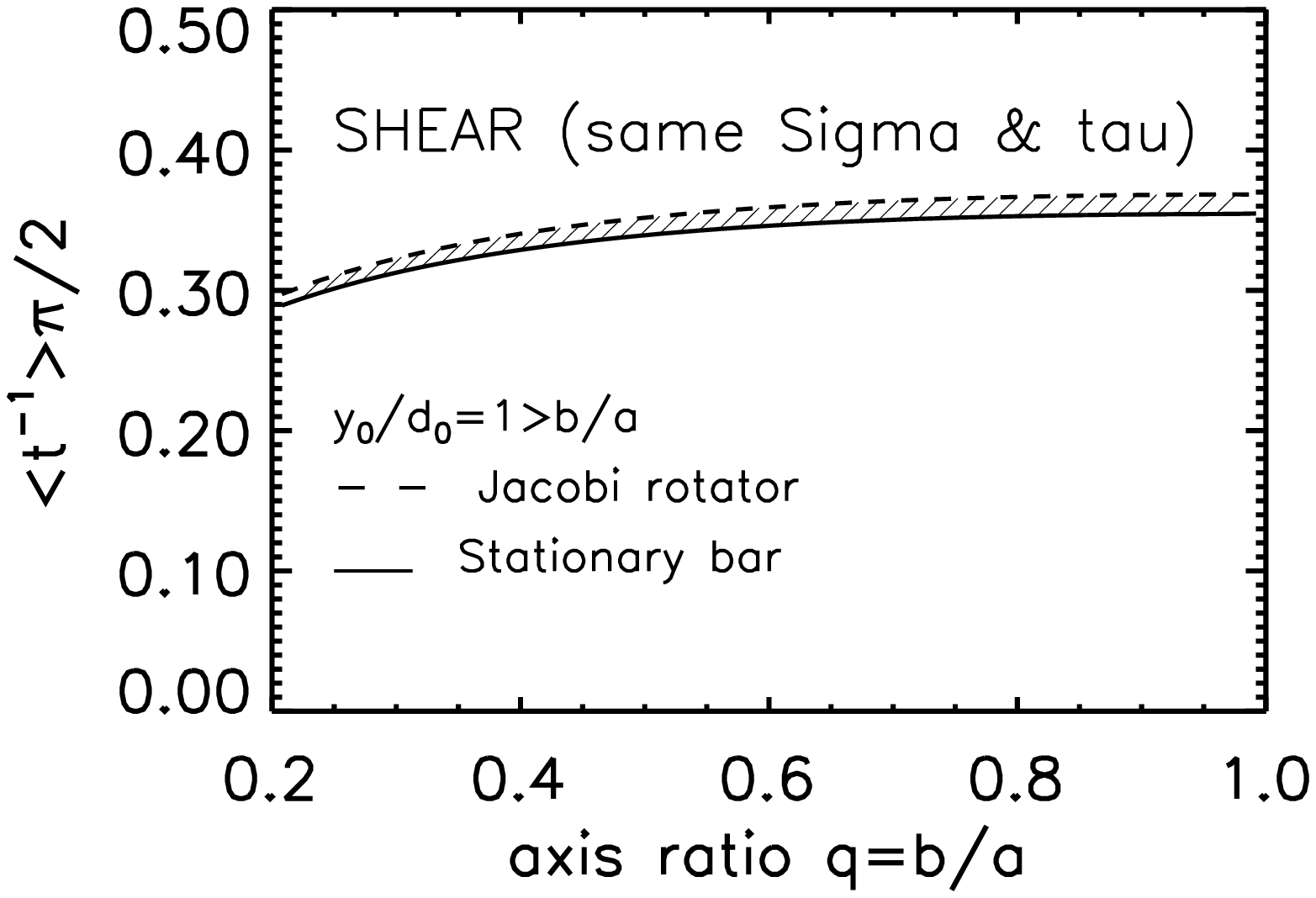}}
\epsfysize=6cm 
\centerline{\epsfbox{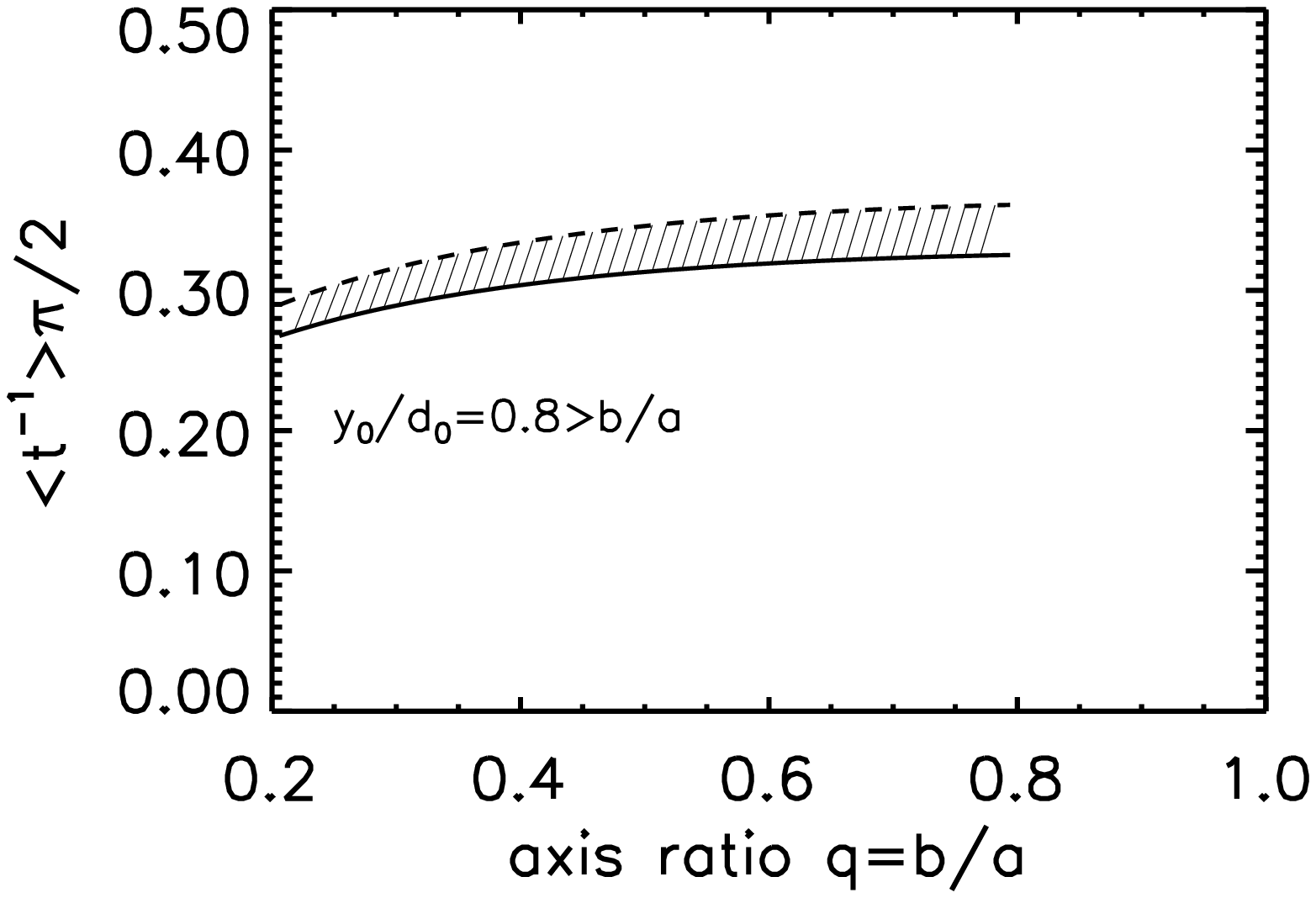}}
\epsfysize=6cm 
\centerline{\epsfbox{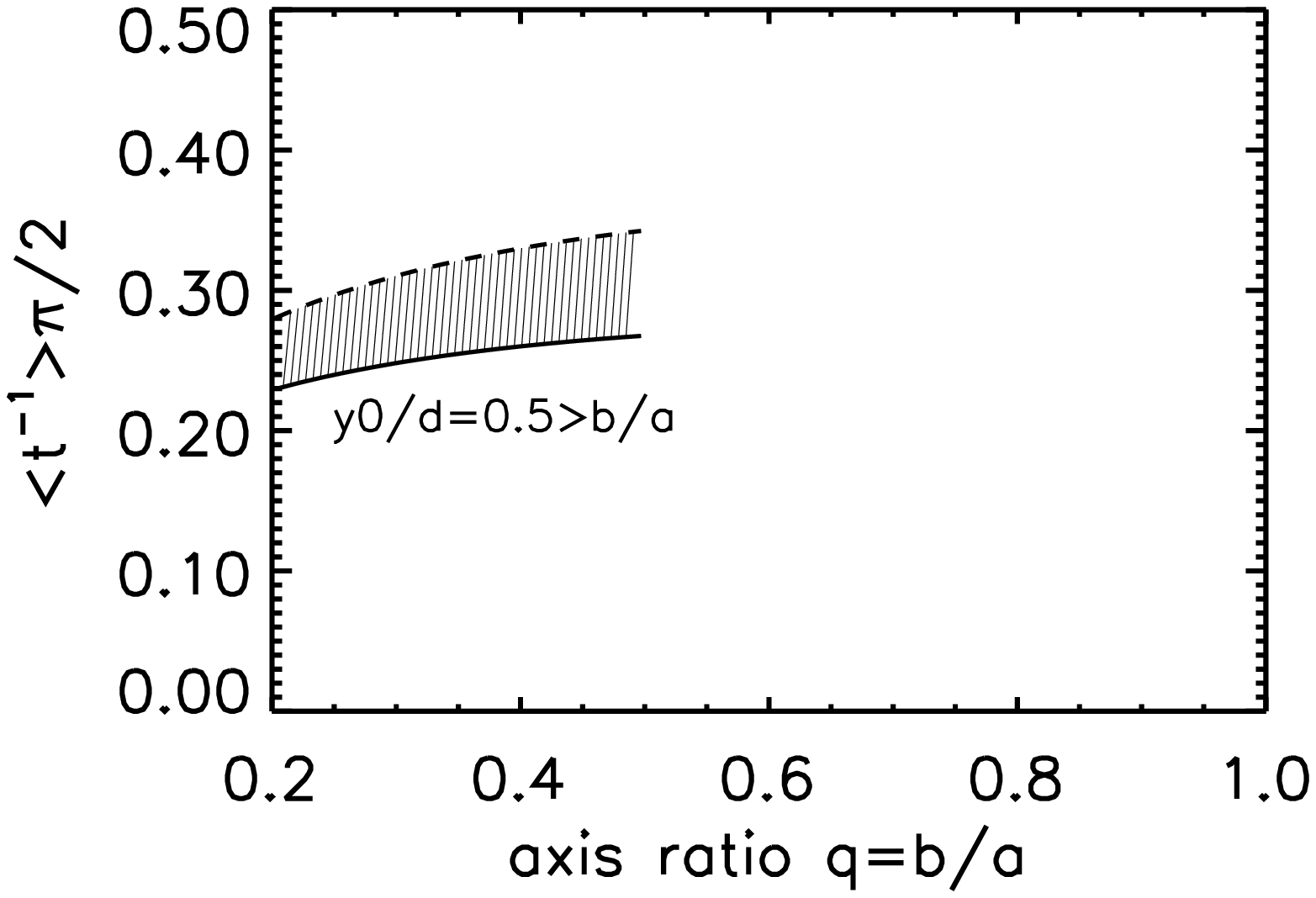}}
\caption{The event rate as a function of the axis ratio for a sequence 
of three bars related by a shear along the line of sight with a fixed
depth to width ratio $d_0 /y_0= 1 /\gamma$. The shaded areas show the
spread by bars with different pattern speed.  The rates are all for a
line of sight through the center of the bar.  The optical depth and
the surface density is invariant in the sequence (cf. 
eq.~\protect{\ref{tauf}}).  }
\label{shear}
\end{figure}

Fixing the bar's volume density and the angle to the bar, the event
rate is generally (at least for models which are bigger in depth than
width) larger for rotating models than for non-rotating models (cf.\
Fig.~\ref{shear} and Fig.~\ref{angle-ratio}).  This is because
rotation increases the relative lens-source speed.  The effect can be
as strong as changing the event rate by $25\%$, as illustrated in
Figure~\ref{shear}).

Since the event distribution is narrower and the rate is higher for
rotating models (Fig.~\ref{profile}), when fitting these models to the
observed event distribution, one would derive a broader mass
distribution with more massive stars as compared to a slightly
narrower spectrum with more low mass objects which would be inferred
if, instead, one adopts non-rotating models (cf.\ eq.~\ref{mv2}).  The
change of the mass spectrum occurs because the observed spread in
$\log t$ is partly due to the spread in lens mass and partly to the
width in $f(\log t)$ of a single lens mass model.

\subsection{Shear}

\begin{figure} 
\epsfysize=6cm 
\centerline{\epsfbox{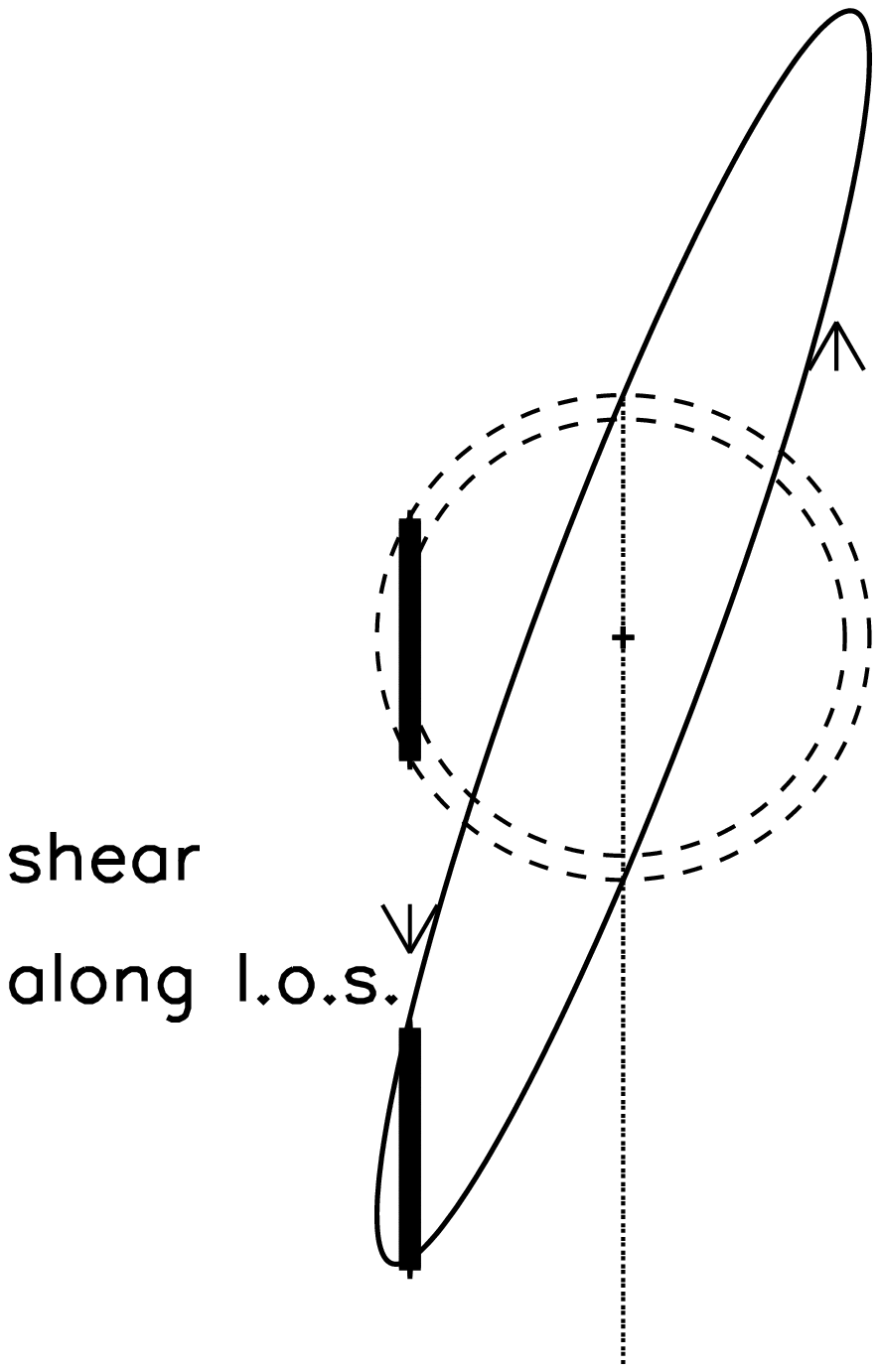}}
\epsfysize=4cm 
\centerline{\epsfbox{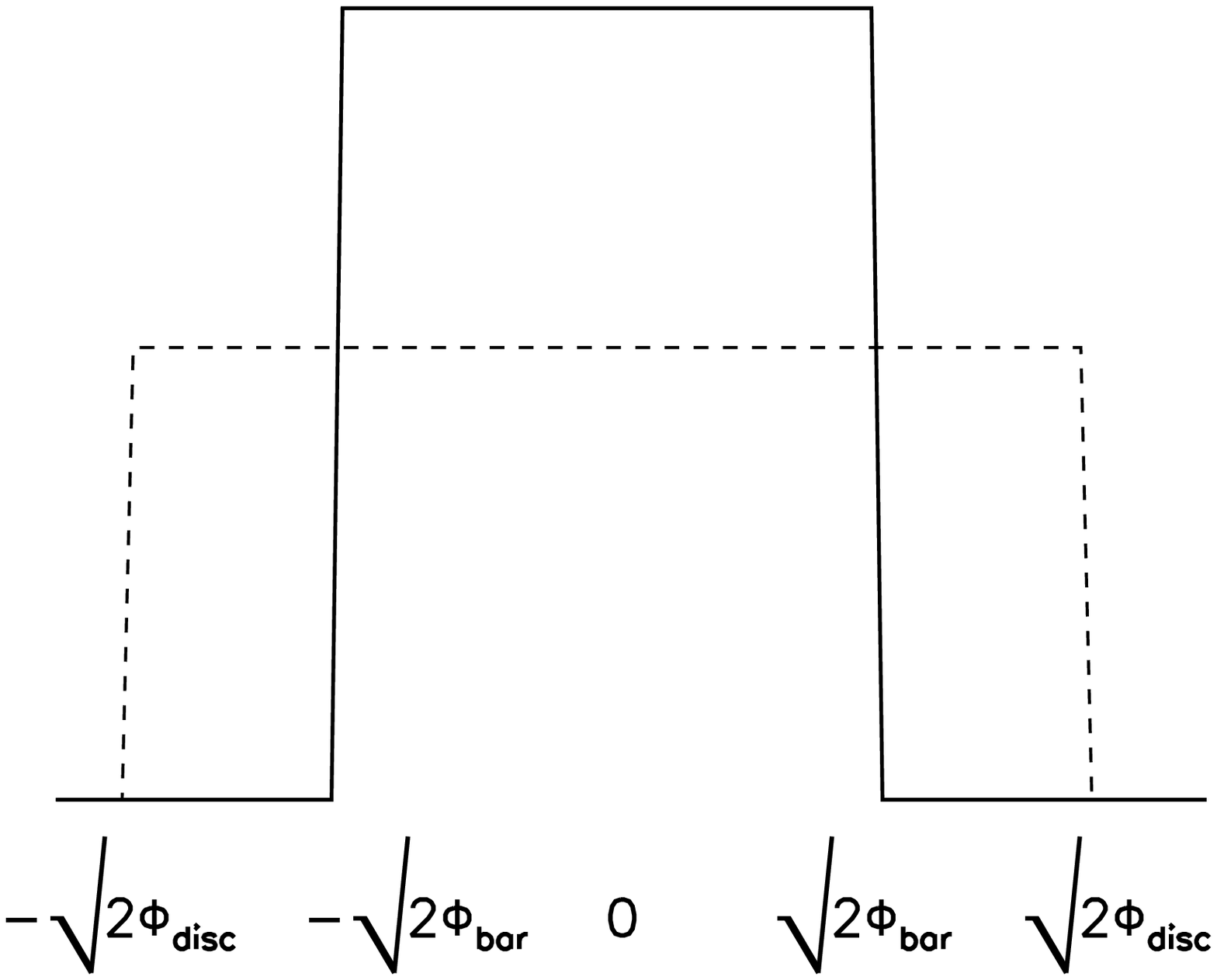}}
\caption{A bar model made by shearing an axisymmetric model along 
the line of sight direction.  For any line of sight, the shear moves
the centroid of the mass distribution without changing the amount of
lens and source, and their relative distance in the line of sight, so
the projected density and the optical depth do not change from model
to model.  The lower panel draws schematic distributions of the
transverse speed at the center (the plus symbol) for a non-rotating
bar or disc.  The elongated bar has a shallower potential and lower
escape velocity $\propto \sqrt{2\Phi_{\rm bar}(0)}$ than the disc
($\propto \sqrt{2\Phi_{\rm disc}(0)}$), where $\Phi$ is the
gravitational potential.  }
\label{physics-shear}
\end{figure}

One can build a sequence of bar models by shearing along the line of
sight (Fig.~\ref{physics-shear}).  Each model of the sequence has the
same surface density and optical depth for an observer located
sufficiently far away from the bar, but all have different event
rates.  At least in the limit of a bar sheared to a needle-shape, it
would have infinitely shallow potential well, hence infinitely small
lens or source velocity, and would predict very few events, but these
would last extremely long.

The sequence of sheared bar models is characterized by a ratio between
the central depth of the bar and the projected size of the bar,
$d_0/y_0$.  The sequence that contains the axisymmetric model
satisfies $d_0/y_0=1$\footnote{For these models $b/a =\tan^2
\alpha$ and the optical depth is the same as for the axisymmetric
model, while bars with maximum optical depth have $b/ a = \tan
\alpha$, and the optical depth is enhanced by $1/\sin 2 \alpha$ 
(ZM).}.  A sequence with $d_0/y_0>1$ produces more optical depth than
an axisymmetric model, while a sequence with $d_0 /y_0 =1 /\gamma < 1$
produces less.  This is because $\tau /\tau^{*} \propto 1/ \gamma =
d_0/y_0$ (cf.\ eqs~\ref{defgamma} and \ref{tauf}).  The observed high
optical depth towards the bulge suggests that $d_0/y_0>1$ for the
Galactic bar, that is, it is longer in depth than across.

Among bars related by a shear transformation, the rate becomes 
smaller when the bar's axis ratio decreases ($b/a \rightarrow
0$), as a result of shallower potential and smaller velocities.  The
rate is largest when the axis ratio is maximum and the bar points
towards or perpendicular to the observer (Fig~\ref{shear}).  The
velocity in a needle-shaped bar is close to zero, and so is the event
rate.  However, excluding such extremely elongated bars which are
likely unstable, we find that for bars with $d_0 /y_0 =1 /\gamma > 1$
the rate is only a weak function of axis ratio; the fractional change is
less than $20\%$.

\section{The rate of extremely short events}

As mentioned in \S 3, the lenses in the bar dominate the events at the
short duration end.  The event distribution is a power-law for small
$t$ (cf.\ eq.~\ref{shortside}).  To quantify the rate of these events,
we consider the accumulative event rate for events shorter than $t$,
\bey
\Gamma(<t) &= &\tau \int_0^t f(\log t) \d (\log t)\\ \nonumber
           &\rightarrow & K t^3 ,~~~t\rightarrow 0,
\eey
where for the Freeman bar the coeficient $K$ is given by 
\beq\label{scalek}
K \equiv {36 \over 35 \pi} \nu^4 \tau.
\eeq
It follows that $\Gamma(<t)|_{t\rightarrow 0}$ is also a power-law of
$t$ with slope 3 (cf.\ Mao \&Paczy\'nski 1996).  The expression of $K$
shows two properties: (a) $\Gamma(<t)|_{t\rightarrow 0}$ is
independent of the rotation parameter $\lambda$.  (b)
$\Gamma(<t)|_{t\rightarrow 0}$ is very sensitive to $\nu$.

It is straightforward to repeat the analysis of \S 4 for the total
rate but now for $K$, since
\beq
{K \over \tau} \propto \nu^4 \propto 
\left({\Gamma \over \tau}\right)^4 \left(1+{\lambda^2 \over 5}\right)^{-4}.
\eeq
For example, 
\beq
{K \over \tau} \propto \nu^4 \propto {1 \over m^2} 
\left({M \over a b}\right)^2
\gamma \left( 1-{y^2 \over y_0^2} \right),
\eeq
where $\gamma (\alpha, {b \over a})$ is given in equation
(\ref{defgamma}).  The general result for extremely short events is
that both $\Gamma(<t)|_{t\rightarrow 0}$ and $K$ are very sensitive
functions of the bar parameters $(M, a, b, \Omega, \alpha)$ and the impact
parameter $y$.

\begin{figure} 
\epsfysize=6cm 
\centerline{\epsfbox{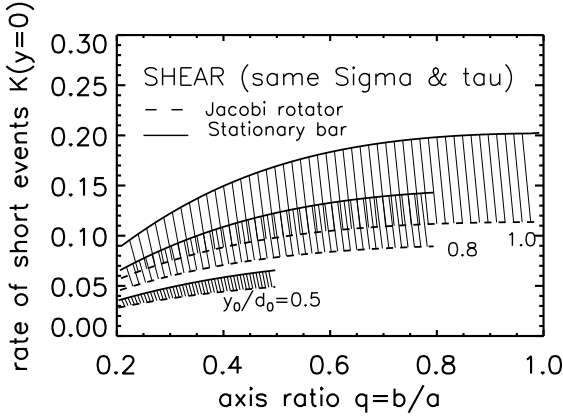}}
\caption{The rate of short events as a function of the axis ratio for 
a sequence of bars related by a shear in the line of sight with a
fixed depth to width ratio $\gamma = {y_0 \over d_0}=1, 0.8, 0.5$ from
the top shaded areas to the bottom.  The shaded areas show the spread
by bars with different pattern speed.  The rates are all for a line of
sight through the center of the bar.  The optical depth and the
surface density is invariant in the sequence.  (cf.\ eq.~\ref{tauf}).}
\label{sheark}
\end{figure}

The value of $K$ can easily vary by a factor of $(1.2-1.3)^4 \approx
2-3$ among models with the same projected density and optical depth
(cf.\ Fig.~\ref{sheark}).  This implies that the mass moment $\left<{1
\over m^2}\right>$ of the lenses in the bar is poorly determined
unless kinematic data are used to break the degeneracy of models with
the same projected density and optical depth.  To clarify this point,
it is helpful to rewrite eq.~(\ref{scalek}) as
\beq\label{m-2}
\left<{1 \over m^2}\right> = {35 \pi \over 36} {G^2 \over c^2}
                             {d_0^2 \over \sigma_0^4 }
     \left[t^{-3}\Gamma(<t)\right]_{t\rightarrow 0, y \rightarrow 0} 
         \propto \sigma_0^{-4}.
\eeq
So the extremely sheared and/or rotating models, which have a lower
$\sigma_0$, give a higher estimate of $\left<{1 \over m^2}\right>$.
This can signficantly influence the determination of the lens mass
spectrum at the faint end (Mao \& Paczy\'nski 1996).

\begin{figure} 
\epsfysize=6cm 
\centerline{\epsfbox{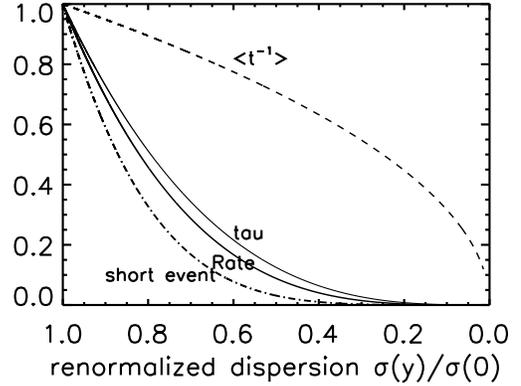}}
\caption{
Dependence of lensing properties on the impact parameter of the line-of-sight.
The horizontal axis is chosen to be the run of velocity dispersion
from the maximum to zero instead of the conventional
run of the impact parameter $y$ from the center outwards 
(but cf. eq.~\protect{\ref{sigmay}} and fig.~\protect{\ref{impact}})
because in this way the curves are likely to be valid also for more 
general bars as well.
The event time scale (dashed line) increases slowly
from the center outward, but the rate of very short events (heavy dot
dashed line) falls off sharply.  The optical depth and the total rate
are in between these two.  }
\label{gradientk}
\end{figure}

\begin{figure} 
\epsfysize=6cm 
\centerline{\epsfbox{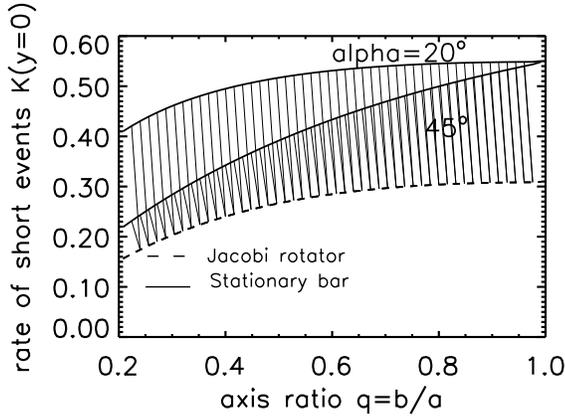}}
\caption{The rate of short events as a function of the axis ratio for a 
sequence of bars with a fixed bar angle.  The shaded areas show the
spread by bars with different pattern speed.  The Jacobi-type rotators
(dashed line) are also isotropic, so $K \propto \sigma_0^4$ is the
same for the two bar angles.  The rates are all for a line of sight
through the center of the bar.  }
\label{ratek}
\end{figure}

$\Gamma(<t)|_{t\rightarrow 0}$ follows the general trend of the total
rate $\Gamma$ with the bar mass and size and the impact parameter but
with a steeper gradient (cf.\ Fig.~\ref{gradientk}).  The dependence
on the bar angle and the axis ratio is somewhat different from the
total rate (cf.\ Fig.~\ref{angle-ratio} and ~\ref{ratek}).
$\Gamma(<t)|_{t\rightarrow 0}$ is largest for an oblate model (cf.\
Fig.~\ref{ratek}), which does not exactly follow the general trend of
the optical depth.

The main difference with the behaviour of $\Gamma$ is that
$\Gamma(<t)|_{t\rightarrow 0}$ decreases with rotation while the total
rate $\Gamma$ increases.  This is because as more kinetic energy is
put into rotation, the dispersion $\sigma_0$ is lowered, so $K \propto
\nu^4 \propto \sigma_0^4$ (cf.\ eq.~[\ref{scaling}]) is lowered.

\section{Conclusions and implications for the Galactic Bar}

We have evaluated the event rate, as well as the event duration
distribution and the optical depth, for a family of bar models known
as the two-dimensional Freeman bars.  We presented several analytical
formulae which show the dependence of the optical depth and event rate
on the bar mass, size, axis ratio, pattern speed, the bar angle and
the impact parameter with a given line of sight.  Models with the same
optical depth and projected density make slightly different
predictions on the event rate and the event duration distribution for
a fixed lens mass.  Here we consider the implications for the bar in
the center of the Galaxy.

\subsection{Event rate as functions of the bar parameters}

Can we reliably generalize the results for the event rate of the
Freeman bars to the Galactic bar?  Realistic bars are three
dimensional and their density structure cannot be modelled by the
Freeman disc.  Furthermore, the orbits in a realistic bar potential
have more variety (e.g., Pfenniger 1984) than in the Freeman bars. For
the Galactic bar, it is most interesting to consider models which can
fit both the COBE map and the observed optical depth.  The Freeman
bars certainly are too simple to model these.

Nevertheless the Freeman bars should be able to offer some insights
and serve as a reference to more complex systems.  The trends found
here for the event rate can be extrapolated with some modifications.
In particular some of the results are due purely to geometry, e.g.,
the event rate and duration as functions of the bar angle.  Some are
generic, e.g., the asymptotic power-laws of $f(\log t)$; the slope at
the short duration side is 3 for both two-dimensional and
three-dimensional models.  Others, e.g., the radial increase of the
event duration, and radial fall-off of the event rate and optical
depth are natural results of a density gradient and the radial
fall-off of velocity dispersion; the gradient of ${\Gamma \over \tau}$
and $\tau$ will depend sensitively on the density gradient.  Because
the event rate is related to the bar dynamics by several integrations
in the lens-source velocity and volume space, which smooth out details
of the phase space, only geometrical or generic effects will show up
in the event rate.  So although true three-dimensional bars have
phase-space distribution functions that may be quite different from
those of the simple Freeman bars, we expect that the latter will
match the basic microlensing physics in the more realistic models.

\begin{figure} 
\epsfysize=10cm 
\centerline{\epsfbox{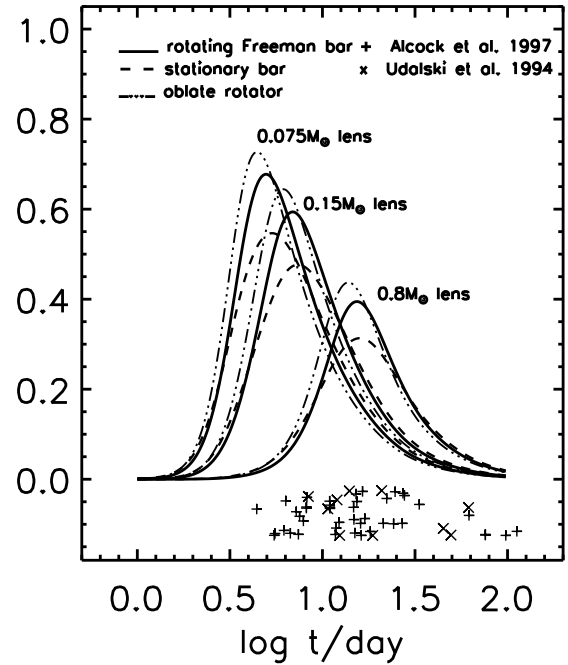}}
\caption{
Distribution of the Einstein radius crossing time of 
observed microlensing events towards the bulge.  Only data
from earlier publications of the MACHO ($+$ symbols) and OGLE
($\times$ symbols) teams are shown.  Also shown are predictions for a line of
sight towards the center from some simple models with the lens mass
being $0.075, 0.15$ or $0.8 M_\odot$ and the detection efficiency
crudely taken into account.  The three underlying dynamical models are
an oblate rotating Kalnajs (1976) disc with semi-axes $a=b=1.5$kpc
($\_$...$\_$), and a simple Freeman (1966) bar pointing $\alpha=30^o$
from our line of sight with semi-axes $(a, b) = (1/\tan\alpha,
\tan\alpha) \times 1.5 = (2.5, 0.8)$kpc in the rotating case
($\_\_\_\_$) and the non-rotating case (- -).  All models have the
same mass $M=2\times 10^{10}M_\odot$, projected size $10^o \sim
1.5$kpc and line-of-sight depth $\sim 1.5$kpc, so to yield almost
indistinguishable maps of the surface density and optical depth. }
\label{compdata}
\end{figure}

\subsection{Uncertainty of predicting the lens mass}

How severe is the non-uniqueness of the lens mass spectrum?  The
common way to derive the mass spectrum in the COBE bar using
microlensing is to compute first an event distribution $f(\log t)$ for
a lens with mass $m$ with a given dynamical model of the bar, and then
to convolve it with such a mass spectrum that the result fits the
observed distribution (e.g., Mao \& Paczy\'nski 1996).  Any
non-uniqueness of the dynamical model propagates to $f(\log t)$, which
in turn propagates to the mass spectrum.

The current best estimates of the lens masses, which use realistic
models for the Galactic bulge or bar, indicate masses near
$0.15M_\odot$, with an uncertain but small $<30\%$ fraction below the
hydrogen burning limit (Zhao, Spergel \& Rich 1995; Zhao, Rich \&
Spergel 1996; Han \& Gould 1996; Han \& Lee 1997).  If these results
were insensitive to the adopted density profile and flattening of the
bar, then a comparable result would be expected from the Freeman bars.
But if one simply scales up the Freeman bar models to the often quoted
mass and size of the Galactic bulge/bar, and makes a straightforward
prediction of the event distribution (cf.\ Fig.~\ref{compdata}), one
finds, quite surprisingly, that the typical lens mass is now boosted
to around $0.8M_\odot$, a factor of $3-5$ times larger than the value
in more realistic models.  Fig.~\ref{compdata} also shows that the
theoretical predictions can not distinguish very well between a model
with $0.075M_\odot$ lenses and a model with the lenses being twice as
massive when different bulge/bar axis ratio, orientation and rotation
speed are considered.  We conclude that the fraction of brown dwarfs
in the Bulge is also sensitive to uncertainties from the dynamical
model.

There are two sources of non-uniqueness of the dynamical model.  The
first is the shear transformation.  Its effect on the distribution
$f(\log t)$ of event durations is generally weak, typically at the
10\% level in terms of the total rate (cf.\ Fig.~\ref{shear}).  Also,
as detailed analysis of the COBE map indicates, the shear
transformation generally leaves detectable signatures in the
left-to-right asymmetry map of the COBE map due to perspective effects
(Binney et al. 1997, Zhao 1997 and references therein).  The second
source of non-uniqueness is the variety of velocity structures that
are possible in the same bar.  This non-uniqueness cannot be
constrained from the COBE map.  As shown also in Fig~\ref{shear}, it
can create typically about $20\%$ difference in the event rate when
one compares a stationary bar with a Jacobi-type rotating bar. This
kind of non-uniqueness is potentially dangerous.

Most relevant to the lens mass function in the bar is the event
duration distribution at the short event side.  In particular, the
mass moment $\left<{1 \over m^2}\right>$ (cf.\ eq.~\ref{m-2}) is very
sensitive to the velocity structure and the shear, and can easily vary
by a factor of $2-3$.  This translates to an uncertainty of the lens
mass (particularly at the lower end) by a factor about $1.4-1.7$.
This implies that any prediction on the faint end of the lens mass
function in the bar is sensitive to details of the adopted dynamical
model of the Galactic bar.

It is possible to reduce the non-uniqueness by fitting the stellar
radial/proper motion velocity of the bar stars.  In fact for the
Freeman bar, it is possible to lift all the degeneracy if one can
measure the line of sight mean streaming motion and velocity
dispersion.  For more realistic three-dimensional models, the
non-uniqueness of the phase space structure is more complicated, and
must be constrained by detailed numerical modeling.  Generally one
would need to fit radial velocities and proper motions at a number of
positions of the bar.  Combined with the COBE map and the microlensing
optical depth, these data will strongly if not uniquely limit the axis
ratio, the angle of the bar and the pattern speed.

\vskip 0.5truecm
\noindent
It is a pleasure to thank Ken Freeman and Simon White for useful
comments, and Dave Syer for help in programming with {\it Mathematica}.
The careful scrutiny by the referee allowed us to improve the
presentation.  HSZ gratefully records the hospitality of the
Max-Planck Institute in Garching where part of this work was done.

\appendix
\section{Orbit boundaries}

In the frame corotating with the bar the curves
bounding the area filled by an orbit of the  Freeman bar
can be written in parametrized
form using eqs (31-33) of F66. We find
\bey
X(\phi) &= &\mu \left[ 1 \pm \left( {\mu^2 \over A^2_\beta} 
+ {k^2_\beta \nu^2 \over k^4_\alpha A^2_\beta} \right)^{-{1 \over 2}} \right]\\
Y(\phi) &= &\nu \left[ 1 \pm \left( {k^4_\alpha \mu^2 
              \over k^4_\beta A^2_\beta} 
+ {\nu^2 \over k^2_\beta A^2_\beta} \right)^{-{1 \over 2}} \right],
\eey
where
the $\pm$ sign indicates the outer and inner boundaries,
\beq
\mu = A_\alpha \cos \phi, \qquad \nu = k_\alpha A_\alpha \sin \phi,
\eeq
and the parameter $\phi$ is an angle ranging between $0$ and $2\pi$.
All quantities have the same meaning as in F66.  

\section{Derivation of the event duration distribution}

The distribution of the relative lens-source speed is
integrated to be (cf.\ eq.~[\ref{fvdef}])
\beq 
\label{FV}
F(V) = \max[0, Q(V)] + \max[0, Q(-V)],
\eeq
where
\bey
4 w_s w_l Q(V) =  &\min&[w_s, w_l+\bar{v}_{ls}-V] \\ \nonumber
                 &+&\min[w_s,w_l-\bar{v}_{ls}+V],
\eey
and $\bar{v}_{ls} \equiv \bar{v}_l-\bar{v}_s$.  Apply the following
change of variables to eq.~(\ref{flogt})
\beq
\tilde{x}_s=d(y) \sin \psi_s, 
\qquad 
\tilde{x}_l=d(y) \sin \psi_l,
\eeq
define $\lambda$ and $\nu$ as in eq.~(\ref{scaling}), and write 
\beq
\delta \equiv \sin \psi_s - \sin \psi_l, ~~~\mu \equiv\nu t,
\eeq
then
\beq
w_{s,l}=\sqrt{3} \sigma(y) \cos \psi_{s,l},
\qquad 
\bar{v}_{s,l}=\lambda \sigma(y) \sin \psi_{s,l},
\eeq
and
\beq
\Sigma_{s,l}={3 M \over 2 \pi a b} (1-{y^2 \over y^2_0}) \cos \psi_{s,l},
\qquad
{V \over \sigma(y) } = 2 \mu^{-1} \sqrt{\delta}.
\eeq
It follows that (cf.\ eq.~\ref{flogt})
\beq
{1 \over \nu} f(\log t) = {2 \ln 10 \over \pi (\nu t)^2 }
                          {I_1(\lambda, \nu t) \over I_2},
\eeq
where
\bey
I_1(\lambda, \mu) &=& \int_{-\pi/2}^{\pi/2} \! \d \psi_s 
                      \int_{-\pi/2}^{\psi_s} \! \d \psi_l \, 
                      \delta \cos^2 \psi_s \cos^2 \psi_l \\ \nonumber
& &\times (\max[0,F_0(\lambda, \mu, \psi_s, \psi_l)] \\ \nonumber
& &  \qquad         +\max[0,F_0(\lambda, -\mu, \psi_s, \psi_l)]),
\eey
and
\beq
I_2 = \int_{-\pi/2}^{\pi/2} \d \psi_s \int_{-\pi/2}^{\psi_s} \d\psi_l \, 
      \delta \cos^2 \psi_s \cos^2 \psi_l = {32 \over 45}.
\eeq
Here we have defined
\bey
F_0 & \equiv & (\nu t) V Q(V) \\ \nonumber
    &=& {\sqrt{\delta} \sqrt{3} \over 6 \cos \psi_s \cos \psi_l } \\ \nonumber
    & & \times ( \min[\cos \psi_s, \cos \psi_l - H_0]  \\ \nonumber
    & & \qquad          +     \min[\cos \psi_s, \cos \psi_l + H_0] ),
\eey
where
\beq
H_0 \equiv {1 \over \sqrt{3}}(\lambda \delta + {2 \sqrt{\delta} \over \mu}).
\eeq
Clearly $I_1$ is generally a dimensionless function of $\lambda$ and
$\nu t$ only, so that 
\beq\label{flogtreduce}
{1 \over \nu} f(\log t) = g_{\lambda}(\nu t) =
         {45  \ln 10 \over 16 \pi (\nu t)^2 } I_1(\lambda, \nu t).
\eeq
The asymptotic relations (cf.\ eqs \ref{shortside} and~\ref{longside})
can be derived by letting $\delta \rightarrow 0$ and $\mu \rightarrow
0$ for $t \rightarrow 0$, and $\mu \rightarrow \infty$ for $t
\rightarrow
\infty$.

{}

\bsp
\label{lastpage}
\end{document}